\documentclass[fleqn,usenatbib]{mnras}

\usepackage{newtxtext,newtxmath}

\usepackage[T1]{fontenc}

\DeclareRobustCommand{\VAN}[3]{#2}
\let\VANthebibliography\thebibliography
\def\thebibliography{\DeclareRobustCommand{\VAN}[3]{##3}\VANthebibliography}

\usepackage{amsmath}
\usepackage{graphicx} 
\usepackage{subcaption}
\usepackage{ulem} 
\usepackage{orcidlink}
\usepackage{CJKutf8} 

\newcommand{\inv}{$^{-1}$}

\title[Hercules: Chemistry]{On the origin of the Hercules group: I. chemical signatures indicating the outer bar origin}

\author[Li, Y. et al.]{
Li, Yusen (李宇森)\orcidlink{0009-0003-8249-6782},$^{1}$ \thanks{E-mail: li.yusen.astr@gmail.com}
Kenneth Freeman\orcidlink{0000-0001-6280-1207},$^{1}$\thanks{E-mail: kenneth.freeman@anu.edu.au}
Helmut Jerjen\orcidlink{0000-0003-4624-9592},$^{1}$
Sven Buder\orcidlink{0000-0002-4031-8553},$^{1,2}$
\newauthor
Michael Hayden\orcidlink{0000-0001-7294-9766},$^{2,3,4,5}$
Ankita Mondal\orcidlink{0009-0006-9737-8609}$^{6}$
\\
$^{1}$Research School of Astronomy and Astrophysics, Australian National University, Canberra, ACT 2611, Australia\\
$^{2}$ARC Centre of Excellence for All Sky Astrophysics in 3 Dimensions (ASTRO 3D), Australia\\
$^{3}$Homer L. Dodge Department of Physics \& Astronomy, University of Oklahoma, 440 W. Brooks St., Norman, OK 73019, USA\\
$^{4}$School of Physics, University of New South Wales, NSW, 2052, Australia\\
$^{5}$Sydney Institute for Astronomy, School of Physics, A28, The University of Sydney, NSW 2006, Australia\\
$^{6}$Centre for Astrophysics and Supercomputing, Swinburne University of Technology, Hawthorn, Victoria 3122, Australia
}

\date{Accepted XXX. Received YYY; in original form ZZZ}

\pubyear{2025}

\begin{document}
\label{firstpage}
\pagerange{\pageref{firstpage}--\pageref{lastpage}}

\begin{CJK*}{UTF8}{gkai}
\maketitle
\end{CJK*}

\begin{abstract}
The Hercules kinematic group is a kinematic anomaly of stars observed in the solar neighbourhood (SNd). In this series of papers, we present a comprehensive study of this structure. This paper focuses on its chemical signatures over several groups of elements. The next paper discusses its kinematical properties. While studies suggested a non-native origin of Hercules stars due to the distinct chemical and kinematic features, previous studies focussed mainly on the Fe abundances. We adopt chemical data with abundances of elements from GALAH and APOGEE to seek further chemical implications on the origin. Our analysis reveals that the low alpha population of the low angular momentum Hercules group is significantly enhanced in iron-peak (Fe, Ni, Mn) and Odd-Z (Na, Al) elements, and slightly deficient in alpha elements (O, Ca, Ti) compared to kinematically local stars. The super enhancement in iron-peak elements and deficiency in alpha elements support their origin from the outer thin bar in the inner Galaxy. Moreover, the enhancement in Na and Al indicates these stars as the youngest stars in the old sequence from the inner thick disc. Hence, the origin of these stars can be related to the outer bar region. These chemical signatures require the underlying dynamical mechanism that forms the Hercules group to be capable of transporting stars in the inner Galaxy out to the SNd. The next paper will consider the Trojan orbits as the favoured mechanism.

\end{abstract}

\section{Introduction}\label{sec:intro}

\begin{figure*}
    \centering
    \includegraphics[width=\linewidth]{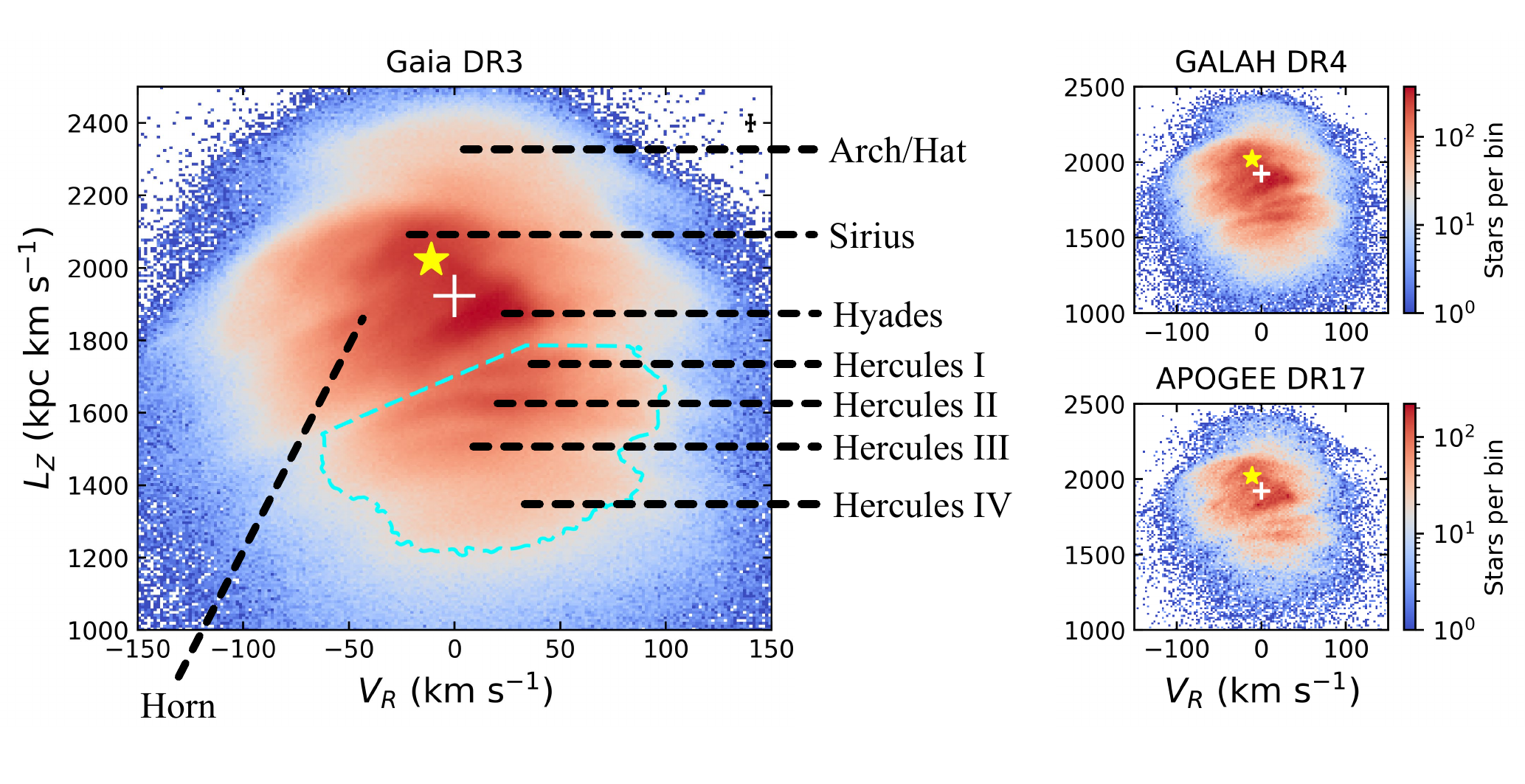}
    \caption{The distribution of stars in the SNd in three surveys in the planar angular momentum-radial velocity plane. The kinematic groups appear as overdensity structures in the plane, consistently throughout three surveys. \textit{Left}: 8.9 million \textit{Gaia} stars in the SNd. The yellow asterisk and the white cross mark the solar and the LSR kinematics, respectively. The Hercules structure, subdivided into four subgroups, is enclosed in the dashed cyan contour and the arbitrary line across the valley below the LSR, including four Hercules subgroups, Hercules I to IV. The black error cross on the top right corner represents the median uncertainties $\epsilon_{V_R} = 1.8$\,km\,s\inv and $\epsilon_{L_Z} = 22$\,kpc\,km\,s\inv. \textit{Upper right}: 391 thousand GALAH stars in the SNd. \textit{Lower right}: 165 thousand APOGEE stars in the SNd.}
    \label{fig:gaialzvr}
\end{figure*}

In the vicinity of the Sun ($ d < 1\,\text{kpc}$), a kinematically asymmetric pattern of stars has been observed in the angular momentum - radial velocity ($L_Z$-$V_R$) plane. This structure is called Hercules, also known as the Hercules stream or the Hercules moving group \citep{PerezVillegas2017ApJ...840L...2P, DOnghia2020ApJ...890..117D}. As marked in Fig.\,\ref{fig:gaialzvr} by the cyan contour, stars in the Hercules group are skewed toward outward radial velocity in Galactocentric cylindrical coordinates, with angular momentum $L_Z \sim 1300$\,to\,$1700$\,kpc\,km\,s\inv, lower than that of stars around the Local Standard of Rest (LSR), $L_Z \approx 1900$\,kpc\,km\,s\inv.

The earliest studies of kinematical structures in the solar neighbourhood (SNd) by \cite{Eggen1950ApJ...111...65E, Eggen1958MNRAS.118..154E, Eggen1983AJ.....88..642E, Eggen1996AJ....112.1595E} identified the overdensities as ``superclusters", stars in dissolved clusters that share similar kinematics. If a supercluster migrates into the SNd, it becomes a ``moving group", leading to the term Hercules moving group. Later studies realised the inclusion of stars with different place and time of birth within these moving group and related these overdensities to radial migration due to perturbations from transient spiral arms \citep{Famaey2005A&A...430..165F:stream}. In that scenario, these structures are refered to as ``dynamical streams", leading to the description of the Hercules stream. We will refer these overdensity kinematical structures as the ``kinematic groups", as ``moving groups" or ``dynamical streams" may imply prior interpretation about its formation mechanism.

In addition to its kinematic signature, studies have identified the distinct chemical properties of the Hercules structure. \cite{Raboud1998A&A...335L..61R} and \cite{Bovy2010ApJ...717..617B:chem} found Hercules stars more metal-enhanced than the local Galactic disc. More recently, \cite{Quillen2018MNRAS.478..228Q} observed an increase in the significance of the Hercules group with [Fe/H], and studies like \cite{Hattori2019MNRAS.484.4540H}, \cite{Khoperskov2022A&A...663A..38K}, and \cite{Liang2023ApJ...956..146L} supported this view using data from \textit{Gaia} \citep{Gaia2023A&A...674A...1G}, APOGEE \citep{APOGEE2022ApJS..259...35A}, LAMOST \citep{LAMOST2012RAA....12.1197C}, and GALAH \citep{Buder2021MNRAS.506..150B_GALAH}. As the metallicity of the interstellar medium around the Sun is near-solar or slightly subsolar \citep{Pagel1981ARA&A..19...77P,Nieva2012A&A...539A.143N:Bstar}, these distinctly different chemical and kinematic properties suggest the origin of Hercules stars outside the SNd. They are believed to have moved from other parts of the Galaxy and preserved the chemical properties from their native environment \citep{Dehnen1998AJ....115.2384D}. 

From another perspective, the asymmetry in $V_R$ inspired studies to dynamically relate the Hercules group to non-axisymmetric structures in the Milky Way. \cite{Kalnajs1991dodg.conf..323K} and \cite{Dehnen2000AJ....119..800D} associtated the Hercules group with the Outer Lindblad Resonance (OLR) of the central Galactic bar \citep[see][ for more details]{Antoja2014A&A...563A..60A,Monari2017MNRAS.466L.113M,Fragkoudi2019MNRAS.488.3324F}. However, more recent models of the inner Galaxy indicates a slowly rotating long bar \citep{Wegg2015MNRAS.450.4050W, Portail2017MNRAS.465.1621P} which disfavours the OLR scenario and argue for other possible mechanisms, including effects from spiral arms \citep{Hunt2018MNRAS.481.3794H_Transient, Michtchenko2018ApJ...863L..37M, Barros2020ApJ...888...75B} or corotation resonance of the bar \citep{PerezVillegas2017ApJ...840L...2P, DOnghia2020ApJ...890..117D, Binney2020MNRAS.495..895B}.

In this series of papers, we present a comprehensive study on features and the formation of the Hercules group. While initial studies focussed on Fe abundances, a standard indicator of metallicity, with the availability of more elemental abundance data from GALAH DR4 \citep{Buder2024arXiv240919858B}, features of more chemical elements of the Hercules group can be studied to obtain a more comprehensive picture. In this first paper we investigate the wider chemical footprint of the Hercules group to gain further insights into the formation and evolution history of these stellar populations. In the second paper \citep{LYS2024arXiv241119097L:II}, we will adopt an idealised barred potential model of Milky Way to investigate the dynamical formation mechanism.

In \S\,\ref{sec:method}, we present our methods for extracting chemical and kinematical data from the surveys. In \S\,\ref{sec:results}, we present the kinematics data from \textit{Gaia} DR3 to identify and define kinematic groups in the SNd, and then look into chemical footprints in the $L_Z$-$V_R$ kinematics plane with GALAH DR4 and APOGEE DR17. Finally, we discuss implications from chemical signatures on the Hercules origin in \S\,\ref{sec:discussion} and summarise our findings and conclusions in \S\,\ref{sec:summary}.

\section{Data}\label{sec:method}

\begin{figure*}
    \centering
    \includegraphics[width=\linewidth]{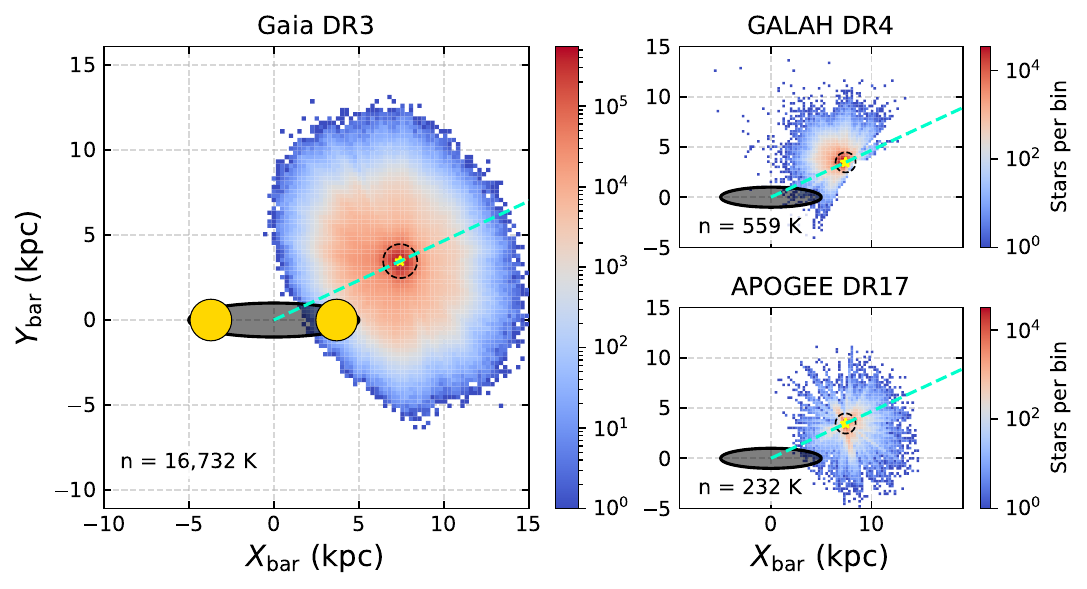}
    \caption{Distribution of stars with Galactic $|z| < 1$ kpc in the surveys, projected onto the Galactic plane in Cartesian $X$-$Y$ coordinates. The Sun and the solar neighbourhood are located $25^\circ$ above the bar's major axis, marked by the yellow asterisk and the dashed black circle. The Galactic bar is marked by the black, semi-transparent ellipse in the centre. Left: 16.7 million stars in \textit{Gaia} DR3, and the super-thin outer bar is marked by the yellow circles; Upper right: 559 thousand stars in GALAH DR4; Lower right: 232 thousand stars in APOGEE DR17. The Logarithmic colour scheme mark the frequency of stars in each bin. The dashed cyan line marks $\phi_{\text{bar}} = 25^\circ$ from the Galactic centre to the Sun.}
    \label{fig:gaiaframe}
\end{figure*}

To minimise the effects of outliers on our results and catch systematic trends, we need to work with a large sample of stars. We make use of \textit{Gaia} DR3 \citep{Gaia2023A&A...674A...1G} which provides accurate proper motions and parallaxes for over 16 million stars after applying the uncertainty restrictions on the entire data set and over 8.8 million stars in the SNd, defined as a cylinder with radius $R_{\text{SNd}} = 1$\,kpc and height $Z_\text{SNd} = 2$\,kpc centred radially and vertically at the sun. For chemical abundances, we are aware of the systematic difference between various surveys due to differences in selection criteria and methods used to infer abundances. Hence, we use abundances from the Apache Point Observatory Galactic Evolution Experiment \citep[APOGEE DR17;][]{APOGEE2022ApJS..259...35A} and the Galactic Archaeology with HERMES survey \citep[GALAH DR4;][]{Buder2024arXiv240919858B} 
to avoid any biases present in a single survey. APOGEE works in the infrared and contains a significant portion of giant stars that can be observed out to relatively large distances from the Sun. On the other hand, GALAH observes at optical wavelengths and focuses on rather local ($< 2$\,kpc) main sequence turn-off (MSTO) stars, from which more accurate age information can be derived. 

To obtain Galactocentric kinematical information in the surveys, we use the \textit{Gaia} parameters ICRS RA ($\alpha$), DEC ($\delta$), PMRA ($\mu_\alpha$), PMDEC ($\mu_\delta$), line-of-sight radial velocity, and stellar parallaxes ($\varpi$) for distances. We adopt the solar location at $R_{\text{GC}} = 8.2$\,kpc away from the Galactic Center \citep{GRAVITY2019A&A...625L..10G, GRAVITY2021A&A...647A..59G,Leung2023MNRAS.519..948L}, with an angular position $\phi_{\text{bar}} = 25^\circ$ above the bar major axis and vertical height at $z_{\text{bar}} = 25$\,pc above the Galactic plane (see Fig.\,\ref{fig:gaiaframe}). Moreover, we adopt the
apparent proper motion of Sgr A* relative to J1745–283 from \cite{Reid2004ApJ...616..872R} as the Galactocentric total angular motion of the Sun, $\text{pm}_l=6.379$\,mas\,yr\inv, and solar peculiar motion $(U, V, W)= (11.1, 12.24, 7.25)$\,km\,s\inv from \cite{Schonrich2010MNRAS.403.1829S}. These parameters assume a Local Standard of Rest (LSR) with a tangential velocity of $v_{T,\text{LSR}}= 235$\,km\,s\inv. Using these assumptions and \textit{Gaia} measurements, we can transform the observed kinematics to Galactocentric cylindrical coordinates, with positions $(R, Z, \phi)$ and velocities $(V_R, V_Z, V_\phi)$ to obtain the planar angular momentum-radial velocity $(L_Z$-$V_R)$ map. We avoid referring to the line-of-sight velocity as ``radial velocity" to avoid confusion. We further use a Galactocentric Cartesian coordinate that aligns the $x$-axis with the bar major axis $(x_{\text{bar}},y_{\text{bar}},z_{\text{bar}})$. Fig.\,\ref{fig:gaiaframe} shows the distribution of \textit{Gaia} stars projected onto the Galactic plane. In this reference frame, the bar (marked by the black ellipse) rotates clockwise and the the SNd is located at the top right relative to the bar, within the black dashed circle. The thin outer bar discovered by \cite{Wegg2015MNRAS.450.4050W} is marked by the yellow circles in the left panel.

To obtain reliable kinematics in the SNd, we restrict our sample to stars with relative errors in parallaxes $\epsilon_\varpi < 10$\,\,per\,cent, proper motions $\epsilon_{\mu_\alpha},\, \epsilon_{\mu_\delta} < 5$\,per\,cent, and absolute error in line-of-sight radial velocity $\epsilon_{\text{RV}} < 5$\,km\,s\inv in addition to positive parallaxes and possibilities to be a single star based on \textit{Gaia} measurements $0.75 <$ \texttt{classprob\_dsc\_combmod\_star}. We note that a systematic bias in parallax measurments exists in \textit{Gaia} DR3 on the order of $ 20-30$\,$\mu$as \citep{Lindegren2021A&A...649A...2L_pi_bias, Gaia2023A&A...674A...1G}. However, as our study focuses on the solar neighbourhood, most stars have parallaxes greater than 1000 $\mu$as. This gives distance errors of less than $3\text{\,per\,cent}$. This error has a limited impact on our results, and we hence neglect the correction for this systematic bias in this work and obtain the distances by directly inverting the parallaxes. We repeat these processes to obtain kinematics for APOGEE and GALAH stars. Note that while the GALAH and APOGEE surveys provide value-added catalogues (VAC) that contain distance and dynamics data, we adopt \textit{Gaia} kinematics for stars in all surveys for consistency.

To obtain the chemical information of the SNd stars, we collect reliable abundances from APOGEE and GALAH. From APOGEE DR17, we obtain abundances for Carbon (C), Oxygen (O), Magnesium (Mg), Aluminium (Al), Silicon (Si), Potassium (K), Manganese (Mn), Iron (Fe), and Nickel (Ni), which are reliably measured in both dwarfs and giants in APOGEE. From GALAH DR4, we consider elements that have more than $50\text{\,per\,cent}$ of all GALAH stars measured with good quality ($\texttt{flag\_X\_Y}=0$). This leaves us with 14 elements: Lithium (Li), Oxygen (O), Sodium (Na), Magnesium (Mg), Aluminium (Al), Silicon (Si), Calcium (Ca), Titanium (Ti), Vanadium (V), Manganese (Mn), Iron (Fe), Nickel (Ni), Yttrium (Y), and Barium (Ba). We also include Europium (Eu) as a tracer of the rapid neutron-capture process (r-process) elements. In GALAH, K is avoided due to an anomaly found in low $|V_R|$, likely caused by the contamination of interstellar potassium \citep{Buder2024arXiv240919858B}. Among the abundance indicators, [X/Fe] is used for elements except [Fe/H] for Fe and A(Li) for Li. Due to the unique formation and depletion history of Li, the GALAH survey prefers 3D NLTE A(Li) as the abundance indicator for Li \citep{EWang2024MNRAS.528.5394W_Li}. We also include the isochrone ages calculated by GALAH.

In all abundances, we apply quality flags ($\texttt{flag\_X\_Y} = 0$ for [X/Y]) in Fe and Mg. Other abundance flags are applied wherever the corresponding abundances are used. For APOGEE stars, we applied the line-of-sight radial velocity flag ($\texttt{RV\_FLAG}=0$) and \textit{Gaia} kinematics error limits same as the \textit{Gaia} sample. For GALAH stars, the overall spectroscopic quality flag is applied ($\texttt{flag\_sp}= 0$) in addition. As GALAH obtain stellar ages by stellar isochrones, and that this method is mostly accurate for main sequence turn-off (MSTO) stars, we restrict the sample to MSTO stars when stellar ages are included.

We notice a difference in the trend of abundances [X/Fe] over [Fe/H] between dwarfs and giants in both surveys. This phenomenon is particularly prominent for Al and Si in APOGEE, in which a ``fish-tail" structure is found in the [Al/Fe]-[Fe/H] and [Si/Fe]-[Fe/H] plane. This could be related to calibration issues or the methods used to infer abundances for different types of stars. However, as this split structure still preserves the general abundance trend, no further corrections were applied.

\section{Analysis} \label{sec:results}

In this section, we consider chemical signatures of stars to seek implications of the origin of stellar populations in Hercules. We first define the kinematic groups in the $L_Z$-$V_R$ plane in the SNd in \S\,\ref{sec:kinematics}. In \S\,\ref{sec:k-group}, we analyse the [Mg/Fe]-[Fe/H] diagram of stars in the SNd kinematic groups to examine the distribution of thin and thick disc populations in different groups. We then conclude the presence of a significant thick disc component in low $L_Z$ Hercules subgroups. In \S\,\ref{sec:X-k}, we colour-code the $L_Z$-$V_R$ kinematic space with chemical abundances and stellar age to seek special properties of Hercules in elements other than Fe and Mg. In \S\,\ref{sec:x-cut}, the abundances are partitioned into bins, in which the distribution of stars in $L_Z$-$V_R$ are investigated to conclude that the high-alpha, low-Fe thick disc stars distribute near flatly in the $L_Z$-$V_R$ plane, and hence contribute little to the kinematic group structures (Hercules, Hyades, Sirius etc.). Finally, we remove the high-alpha population in \S\,\ref{sec:x-k-td} and plot the distribution of abundances and age in the $L_Z$-$V_R$ kinematic space to observe a super metal-rich patch of stars at the low $L_Z$ Hercules subgroups III \& IV that has chemical signatures of the inner Galaxy.

\subsection{Kinematic groups in the $\mathbf{L_Z}$-$\mathbf{V_R}$ plane}
\label{sec:kinematics}

We first have a look at the kinematics of stars in the solar neighbourhood (SNd) defined by the cylindrical volume defined by Galactocentric cylindrical cartesian coordinates $(x-x_\odot)^2 + (y - y_\odot)^2 < 1\,\text{kpc}^2$ and $|z| < 1$\,kpc. This defined the distinct kinematic groups in the SNd including Hercules.

In Fig.\,\ref{fig:gaialzvr} we present the distribution of these stars in the angular momentum ($L_Z = V_T R$) versus radial velocity ($V_R$) space, where $V_T$ and $V_R$ are the Galactocentric tangential and radial velocities in the non-rotating cylindrical system. In each of the three panels, the solar kinematics is marked by the yellow asterisk and the LSR kinematics by the white cross. In the \textit{Gaia} DR3 panel, we observe and define kinematic groups with the help of contour lines. The majority of overdensities around the LSR can be considered as kinematically local. These structures are separated with other structures with lower $L_Z$ by an under-density valley.

Below the valley, the Hercules group is the over-density stripes highlighted by the dashed cyan line. Stars in the Hercules group cover a significant portion of that kinematic space which can be further subdivided into four distinct substructures: the most prominent substructures Hercules I and II are the two horizontal stripes asymmetric in $V_R$ with angular momenta ranging from 1550\,to\,1750\,kpc\,km\,s\inv. A less asymmetric and fainter third subgroup Hercules III is found at $L_Z \sim 1450$\,kpc\,km\,s\inv and an even fainter, more asymmetric fourth over-density (Hercules IV) is observed with the lowest angular momentum $L_Z \sim 1300$\,kpc\,km\,s\inv. Among these substructures, Hercules I, II, and IV are heavily skewed towards positive radial velocities $V_R$, moving away from the Galactic centre, while Hercules III is more symmetric in $V_R$.

We also identify four other well-known kinematic groups above the dashed cyan line. From high to low $L_z$ values there is the symmetric Arch/hat ($L_Z > 2200$\,kpc\,km\,s\inv), the left-biased solar-habiting Sirius group, the most populated Hyades group, and the Horn on the left of Hyades. We refer to the structures in the higher fraction of the kinematics plot consistently with previous studies \citep{Monari2019A&A...626A..41M, Bernet2022A&A...667A.116B, Khoperskov2022A&A...663A..38K}. While these higher $L_Z$ structures are not the primary focus of this paper, their origins 
have been related to bar or spiral resonances, but are not yet well-understood. Note that in comparison to other studies, we adopt a broader definition of the Hercules group. In some works, Hercules III is considered as HR 1614 \citep{Kushniruk2020A&A...638A.154K}, and Hercules IV is named the Arcturus group in \cite{Bernet2022A&A...667A.116B}.

The graphs in Fig.\,\ref{fig:gaialzvr} show that the kinematic groups are consistently found in the three surveys. A noticeable ``valley" is observed between the low $L_Z$ Hercules and high $L_Z$, more kinematically local groups. The Hercules group makes up $23\text{\,per\,cent}$ of all the SNd stars in \textit{Gaia} DR3, $27\text{\,per\,cent}$ in GALAH, and $21\text{\,per\,cent}$ in APOGEE, respectively. The Hercules structures exist regardless of the selection criteria of surveys and hence cannot be the result of a selection bias in the surveys. 

\subsection{Thin and thick disc contribution in kinematic groups}\label{sec:k-group}

\begin{figure}
    \centering
    \includegraphics[width=\linewidth]{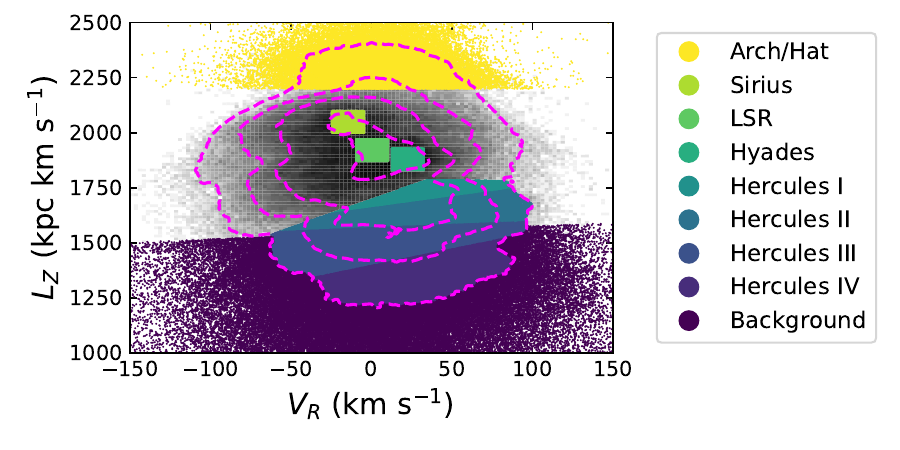}
    \caption{The selection of stars into kinematic group samples in the local $L_Z$-$V_R$ kinematic space. The dashed magenta contour lines show density structures in Fig. \ref{fig:gaialzvr} and can therefore function as a reference to the position of kinematic groups in the $L_Z$-$V_R$ plane. It compares density structures with chemical structures in \S\,\ref{sec:X-k} and \S\,\ref{sec:x-k-td}. The selected coloured regions are used in the following sections to select stars into kinematic group samples.}
    \label{fig:selection}
\end{figure}

It is well known that disc stars in the Milky Way have an old, metal-poor, alpha-rich thick disc population and a metal-rich, alpha-poor thin disc population that covers a broad range of ages \citep{Wallerstein1962ApJS....6..407W, Tinsley1979ApJ...229.1046T, Tinsley1980FCPh....5..287T, Yoshii1982PASJ...34..365Y, Gilmore1983MNRAS.202.1025G}. The two populations are well distinguished in the [Mg/Fe]-[Fe/H] diagram. 
\cite{Bensby2014A&A...562A..71B} and \cite{Hayden2015ApJ...808..132H} employed this diagnostic tool to study the Milky Way disks and observed a negative gradient in Fe to the outer disc. More recent studies completed the picture by finding a positive inner galactic radial metallicity gradient and associating the [Fe/H] peak of the low alpha sequence to the outer flat bar \citep{Wegg2015MNRAS.450.4050W, Wylie2021A&A...653A.143W}.
While the contributions of these two stellar populations in different kinematic groups contain rich information, the chemical difference of these two distinct populations can cause abnormal trends that could obstruct the real pattern. Therefore, before we observe the inconsistencies in \S\,\ref{sec:X-k} and later resolve them in the following sections, in this section, we adopt the [Mg/Fe]-[Fe/H] diagram to understand the difference in thin/thick disc contribution in different groups.

\begin{figure*}
    \centering
    \includegraphics[width=.8\linewidth]{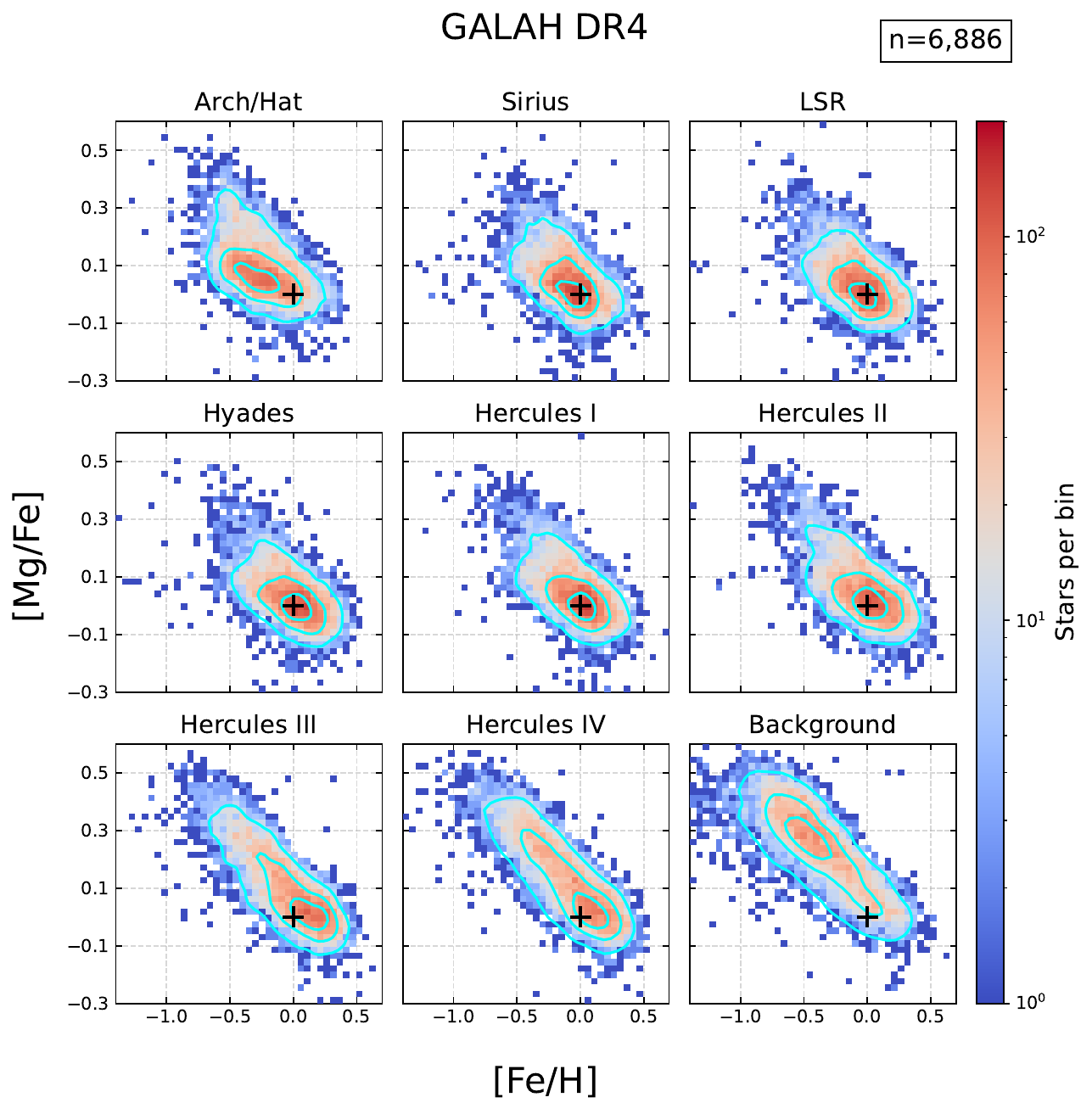}
    \caption{The [Mg/Fe]-[Fe/H] distribution of GALAH stars for the eight kinematic groups in the SNd (see Fig.\,\ref{fig:selection}) and for the background ($L_Z<0.3\cdot V_R+1550$\,kpc\,km\,s\inv and outside Hercules). Kinematic groups with lower $L_Z$ ($L_Z < 1600$\,kpc\,km\,s\inv) are found to contain a larger portion of high alpha thick disc stars. The low alpha thin disc peak of Hercules III and IV are more Fe-enhanced, while that of the Arch/Hat is more deficient than others.  Bin numbers of $40\times40$ are adopted in all 2D histograms. The cross at $(0,0)$ marks the solar abundance for reference. The cyan contours encircle 90, 60, and 25 per\,cent of stars from the peak of the distribution in each graph. The number of stars in each sample is normalised to 6,886 by random removal to ensure a constant sample size for better comparison.}
    \label{fig:mgfe_galah}
\end{figure*}

\begin{figure*}
    \centering
    \includegraphics[width=.8\linewidth]{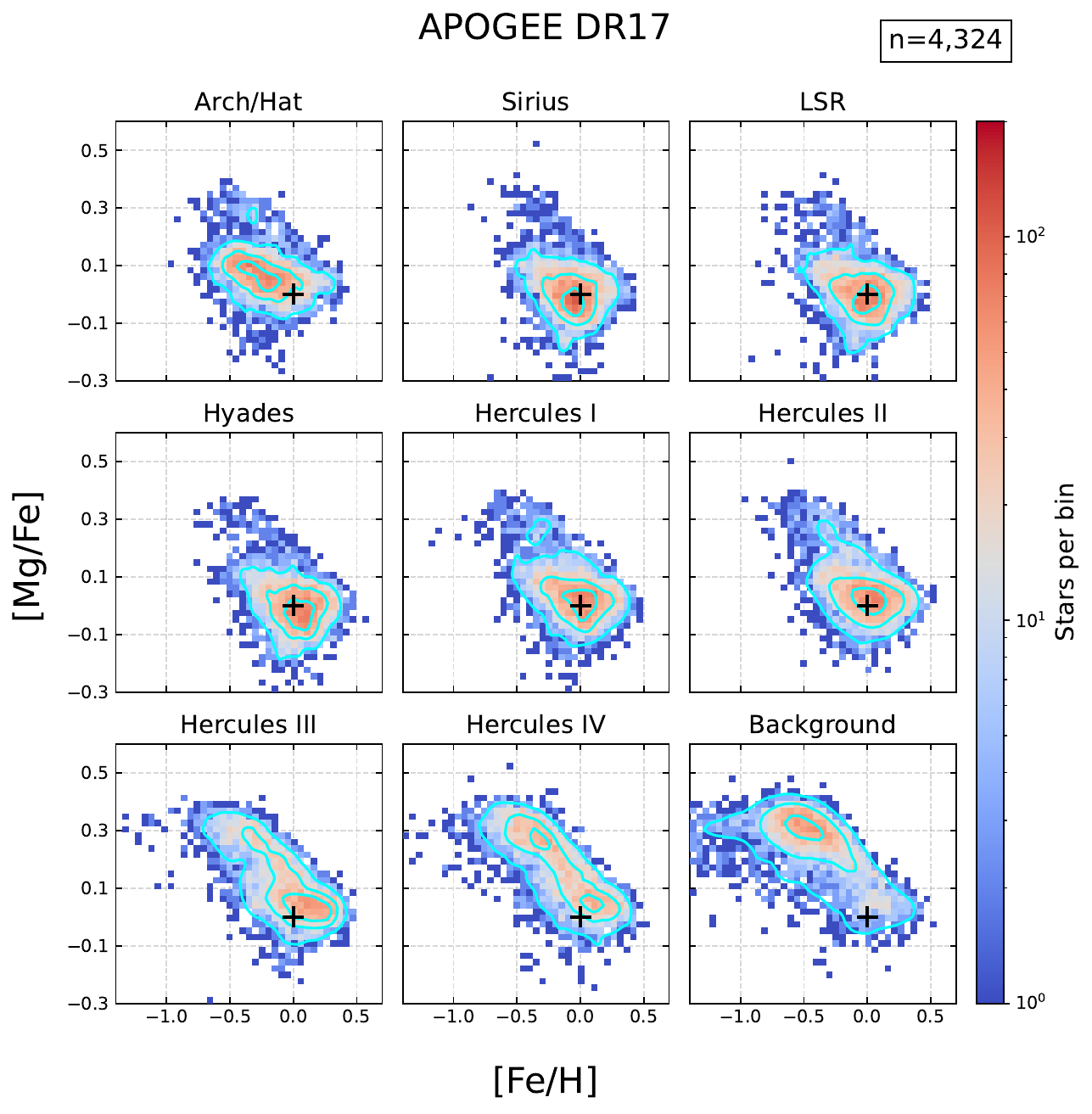}
    \caption{The [Mg/Fe]-[Fe/H] distribution of APOGEE stars in eight samples of the SNd kinematic groups and background. Same as Fig.\,\protect\ref{fig:mgfe_galah} except with APOGEE data. The chemical trends are found to be consistent with GALAH except for a better bimodality separation. The number of stars in each sample is normalised to 4324.}
    \label{fig:mgfe_apogee}
\end{figure*}

From all the kinematic groups defined in Fig.\,\ref{fig:gaialzvr}, we obtain samples of stars as in Fig.\,\ref{fig:selection}. A low angular momentum background sample ($L_Z<0.3\cdot V_R+1550$\,kpc\,km\,s\inv) outside the Hercules cyan contour mark in Fig.\,\ref{fig:gaialzvr} is included in addition. The [Mg/Fe]-[Fe/H] distributions of different samples are presented in Fig.\,\ref{fig:mgfe_galah} for GALAH and Fig.\,\ref{fig:mgfe_apogee} for APOGEE, respectively. The sample size of stars in each subpopulation is normalised by uniform random removal to avoid visual and statistical biases. In each graph, three cyan contours are plotted to include 90 per\,cent, 60 per\,cent and 25 per\,cent of all stars. The solar abundance is marked by the black cross for reference. Bin sizes of 0.05 in [Fe/H] and 0.02 in [Mg/Fe] are adopted in all 2D histograms to ensure a smooth profile while capable of capturing substructures in the [Mg/Fe]-[Fe/H] space. This is also consistent with typical errors presented in [Fe/H] and [Mg/Fe] in Figs.\,\ref{fig:lzvr_X_galah} and \ref{fig:lzvr_X_apogee}.

\begin{table*}
    \centering
    \begin{tabular}{| c ||c|c|c|c|c|c|c|c|c|}
        \hline
        
         Sample      & Arch/Hat  & Sirius    & LSR   & Hyades   & Hercules I & Hercules II & Hercules III  & Hercules IV & Background   \\
         \hline
         \textbf{GALAH}\\

         [Fe/H]     &-0.20 & -0.04 & 0.01 & 0.06 & 0.06 & 0.06 & 0.11 & 0.06 & -0.35 \\
         
         [Mg/Fe]    & 0.06       & 0.02      & 0.00  & 0.00     & 0.00   & 0.02 & 0.02    & 0.04   & 0.26  \\
         \hline
         \textbf{APOGEE}\\

         [Fe/H]     &-0.35 & -0.04 & 0.06 & 0.06 & -0.04 & 0.06 & 0.11 & -0.25 & -0.40 \\
         
         [Mg/Fe]    & 0.11       & 0.00      & 0.00  & -0.05     & 0.04   & 0.04 & 0.04    & 0.26   & 0.30  \\
         \hline
         
    \end{tabular}
    \caption{The peaks in the [Fe/H]-[Mg/Fe] distributions of GALAH and APOGEE samples of kinematics groups as shown in Figs. \ref{fig:mgfe_galah} and \ref{fig:mgfe_apogee}. Note that the distributions of the Hercules IV stars in APOGEE has two local maxima and the values listed represents the thick disc peak.}
    \label{tab:mgfe}
\end{table*}

We see a clear bimodality caused by the thick and thin disc stars in the Galaxy in the background sample of APOGEE with local peaks at $(\text{[Fe/H]},\text{[Mg/Fe]}) = (-0.50,0.30)$ and ($0.1,0.05$). The bimodality is less distinguishable in GALAH, but the thick disc population of high [Mg/Fe], and low [Fe/H] is still prominent in the data. Using this background map as a reference, we observe that the high $L_Z$ groups LSR, Sirius, and Hyades contain very few thick disc stars, while the Arch/hat with the highest $L_Z$ has slightly more. In the Hercules subgroups, the fraction of thick disc stars gradually increases with the decrease in angular momentum. In Hercules III and IV, the metal-poor thick disc population is as abundant as the metal-rich thin disc population. The abundance peak for each group is presented in Table\,\ref{tab:mgfe}.

By focusing on the metal-rich, alpha-poor thin disc population, we observe that the distribution of the Arch stars is shifted strongly to the Fe-poor, Mg-poor end. This is a feature found in outer thin disc stars \citep[e.g.][]{Guiglion2024A&A...682A...9G}. The low alpha population of Sirius, LSR, and Hyades are similar to solar values while the peaks shift slightly towards high-Fe, and low-Mg with the decrease of angular momentum. The distribution of the thin disc population remains nearly identical between Hyades, Hercules I, and Hercules II samples in GALAH while Hercules I and II have slightly higher [Mg/Fe] than Hyades. Below Hercules II, Hercules III and IV groups are found more Fe-enhanced than the other six groups in both GALAH and APOGEE samples, with [Fe/H] $\sim 0.2$ dex. 

This [Fe/H] abundance is significantly higher than most thin disc stars in the SNd and the predicted [Fe/H] for the solar vicinity, around $\sim -0.1$ dex \citep{Lian2023NatAs...7..951L} slightly lower than the solar value. In addition, the interstellar medium (ISM) in the solar neighbourhood is measured to have near-solar or slightly sub-solar  metallicity \citep{Pagel1981ARA&A..19...77P, DeCia2021Natur.597..206D}. These signatures suggest a low alpha population in the Hercules III and IV subgroup, which is different to stars chemically local to the solar vicinity. They are likely to have an origin closer to the Galactic centre, moved outwards, and transported to the SNd through either stochastic radial migration or more stable mechanisms like orbits. Note that some super metal-rich stars ([Fe/H] $> 0.25$) are also found in high $L_Z$ groups. In summary, the low alpha thin disc population in Hercules III and IV is very different to the stars kinematically similar to the LSR. This indicates the origin of these stars to be different from other local stars.

We notice an interesting bimodality in Fig.\,\ref{fig:mgfe_apogee}. In the low alpha sequence of the Arch/hat sample, the major peak is detected at ([Fe/H],[Mg/Fe]) $=(-0.4, 0.1)$ in addition to the more standard local peak at $(-0.25,0.05)$ for the other high $L_Z$ groups. This new peak is less Fe-enhanced and more alpha-enhanced than the other peak. Weaker signals of the same bimodality can be observed in other high $L_Z$ groups as well as Hercules I and II. The bimodality is less pronounced in the full GALAH data, but the structure is uncovered in the main sequence turn-off (MSTO) population. The existence of this second peak in GALAH DR4 may be associated with neural networks employed to extract abundances from spectroscopy \citep[see][]{Buder2024arXiv240919858B}. However, this is uncertain as the same structure is found also in APOGEE. 

\subsection{Abundance distribution in the kinematic space}\label{sec:X-k}

After noticing the Fe, Mg abundance variation in the stellar populations of the different kinematic groups in the SNd, we expect those groups to show differences also in other elements. To test this supposition, we cover the kinematics plane with $30\times30$ hexagonal bins across the $V_R \in (-150,150)$\,km\,s\inv and $L_Z \in (1000,2500)$\,kpc\,km\,s\inv range, resulting in bin sizes of 10\,km\,s\inv in $V_R$ and 50\,kpc\,km\,s\inv in $L_Z$. We colour-code each bin with averaged abundance or stellar age, where red is high and blue is low. To avoid low number statistics, we exclude bins with frequencies less than 10. The bin size is chosen to compromise the noise in data and substructures. In each plot, the number density contours in Fig.\,\ref{fig:selection} are plotted as yellow dashed lines on top of the chemical distribution to help locate the kinematic groups. In this section, we adopt abundance data from GALAH and APOGEE and will show that, while the abundance distribution shows peak or anti-peak structures in most elements, the inconsistency in Fe-peak elements and between two surveys indicates the potential effect of the mixture of the thick and thin disc populations.

\begin{figure*}
    \centering
    \includegraphics[width=\linewidth]{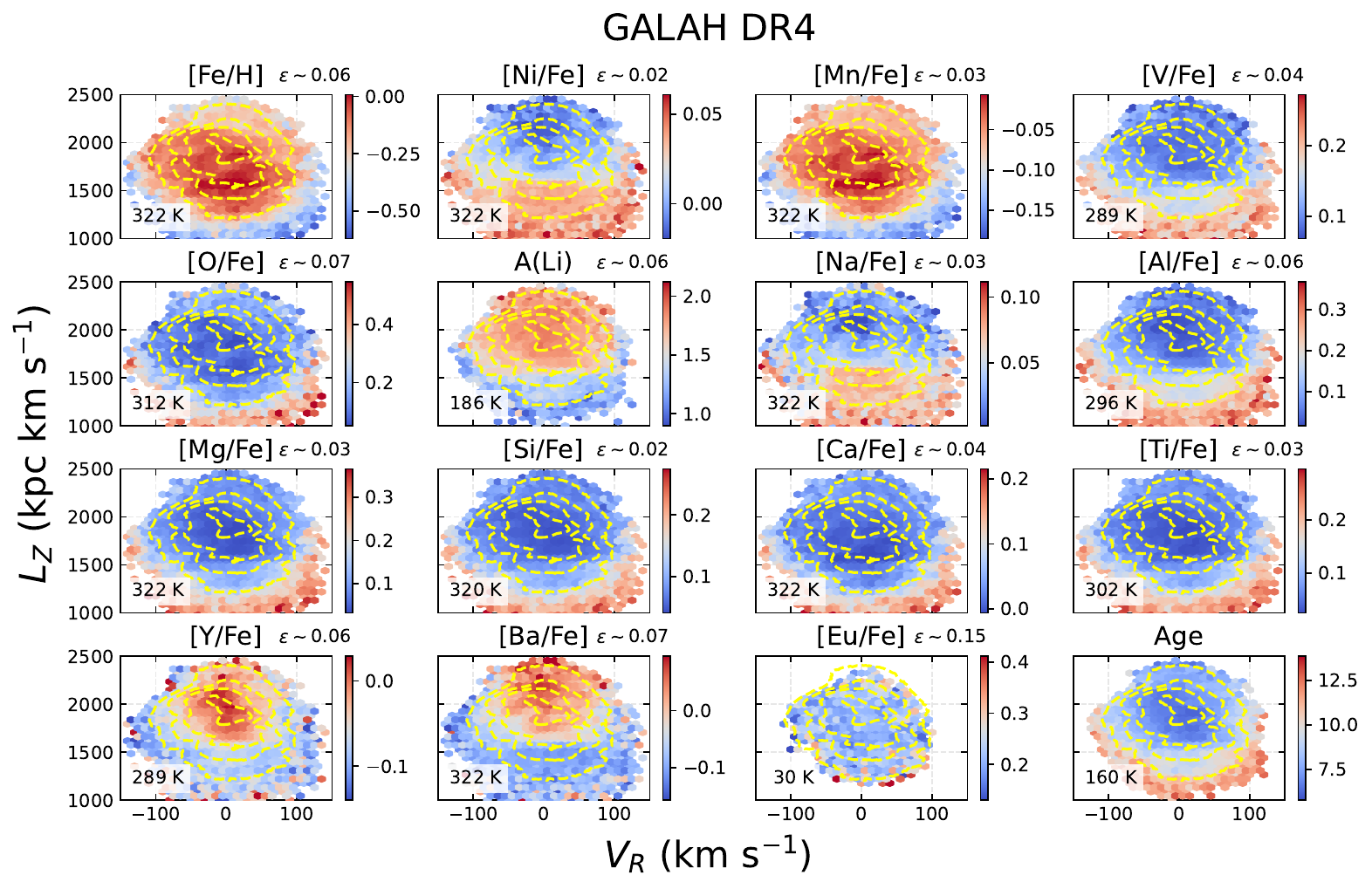}
    \caption{The $L_Z$-$V_R$ kinematics space coloured by 15 reliable element abundances and age for the GALAH DR4 sample. While many kinematic groups appear as peaks or anti-peaks in many abundances, the inconsistency of trends in the Fe-peak elements (top row) challenges the reliability of the analysis. The flagged stellar abundances in each hexagonal bin are averaged and coloured to scale. The density contours in the $L_Z$-$V_R$ plane from Fig.\,\ref{fig:selection} are plotted in yellow dashed lines as the reference to the position kinematic structures. The total number of stars in each plot is given in the bottom left corner. The median uncertainties of chemical abundances reported the GALAH survey $\epsilon$ is given next to the abundance subtitles.}
    \label{fig:lzvr_X_galah}
\end{figure*}

We first present the 14 reliable abundances in GALAH (more than $50\text{\,per\,cent}$ of stars measured with good quality), r-process candidate Eu, and stellar age in Fig.\,\ref{fig:lzvr_X_galah}. These elements can be subdivided into six categories: Fe, Ni, Mn, and V in the first row are iron-peak elements mainly produced by type Ia supernovae. In the second row are the light elements O and Li and the Odd-Z elements Na and Al. The 3rd row shows the classic alpha elements Mg, Si, Ca, and Ti produced by core-collapse supernovae, and in the bottom row are the s-process elements Y and Ba, the r-process element Eu, and stellar age. Note that while O is also considered as an alpha element, we categorise it as light element to keep it consistent with the GALAH survey \citep{Buder2024arXiv240919858B}.

At first glance, we observe that the global trend of abundances mainly varies over angular momentum and that the kinematic groups around LSR generally have significantly different abundances compared to background stars with low $L_Z$ or high $|V_R|$. We consider each category of elements together in their general abundance trend between high $L_Z$ disc stars and low $L_Z$ background stars as well as noticeable trends among the kinematic groups defined in Figs.\,\ref{fig:gaialzvr} and \ref{fig:selection} (Arch/Hat, Sirius, Hyades, Horn, and Hercules I to IV). 

\begin{itemize}
    \item \textbf{Fe-peak elements}: the most typical iron peak elements [Fe/H] and [Mn/Fe] show almost identical distributions: richer at higher $L_Z$ and lower in the background. Ni and V, however, show a different general trend with more enhancement at lower $L_Z$. In both Fe and Mn, we see significant over-enhancement in high $L_Z$ Hercules subgroups, Hyades, and a slightly weaker signal in Horn. In Ni, a sharp increase in enhancement is observed at $L_Z \sim 1600$\,kpc\,km\,s\inv, resulting in a low $L_Z$ patch feature covering Hercules III and IV. This low $L_Z$ patch is a strong feature in Ni and a similar hint can be seen in V. While no obvious structures can be inferred on the top fraction of V, a significant enhancement is observed in Ni for LSR, Sirius, and Hat.

    \item \textbf{Light elements}: Among the two light elements, O shows a global trend of more enhancement in the background similar to V, and Li follows Mn which is more enhanced at high $L_Z$. Slightly, O shows a deficiency in kinematic groups Sirius, Hyades, and Hercules. In particular, Li shows an increase in abundance with the increase of $L_Z$ and is significantly more enhanced at the Hat. In addition, Li also shows a sharp transition into the Li-poor low $L_Z$ patch at Hercules III and IV. This global trend may be an indication of the positive Galactic radial Li gradient. Combined with the age distribution, this is a reflection of the more intrinsic negative Galactic radial age gradient due to the inside-out formation \citep{Chiappini2001ApJ...554.1044C, Bird2013ApJ...773...43B}: high $L_Z$ stars have larger guiding radii ($R_g$ in axisymmetric approximations) and originate from the outer disc; these stars then include lower stellar ages due to the negative radial age gradient; finally, as Li-enhancement is anti-correlated with stellar age \citep{EWang2024MNRAS.528.5394W_Li}, younger stars with higher $L_Z$ are more enhanced in Li.

    \item \textbf{Odd-Z elements}: The two Odd-Z elements, Na and Al, both show a pattern similar to Ni and inverse to Li. The SNd stars are found Odd-Z poor at higher $L_Z$ and Odd-Z rich in the low $L_Z$ background. Similar to Ni, the transition into the low $L_Z$ patch is also observed in Na and Al. This feature is stronger in Na and less prominent in Al. In addition, Sirius is very deficient in both Na and Al.

    \item \textbf{Alpha elements}: On the third row, the four alpha elements show nearly identical patterns. Similar to O, alphas are more enhanced in the background and less enhanced in the main disc population. Among the four, Ca shows some finer structures that show anti-peaks at Hercules I, II, Hyades, and the Horn.

    \item \textbf{S-process elements}: On the last row, Y and Ba are less enhanced at the background and Low $L_Z$ Hercules subgroups. Both s-process elements show significant enhancement in a low $|V_R|$ strip from Hercules II, I, Hyades, LSR, Sirius, to the Hat. The maximum at this s-process-rich strip happens at Sirius. An arguable transition like Ni and Odd-Zs is also observed in Ba resulting in a Ba-poor low $L_Z$ patch.

    \item \textbf{R-process element}: Likely due to the limitation in the quality and scarcity of data, Eu shows no distinguishable pattern in the kinematics space. The distribution of Eu is nearly uniform in the main patch and shows random variations on the edges where more scarce data is present and the outliers dominate the mean. 

    \item \textbf{Age}: In the last figure, the isochrone age provided by GALAH is distributed similarly to Al and alphas. Background stars are typically significantly older than the high $L_Z$ stars in the kinematic groups. Like Ni, Na, and Al, while weaker, the age shows evidence of a minimum in the Sirius group. Moreover, stellar age shows an inverted trend to Li and s-process elements. This is expected due to the strong relation between the stellar age and formation of s-process elements \citep{DOrazi2009ApJ...693L..31D:Ba}, and between age and the depletion of Li \citep{Baumann2010A&A...519A..87B:Li}. 
\end{itemize}

\begin{figure*}
    \centering
    \includegraphics[width=.75\linewidth]{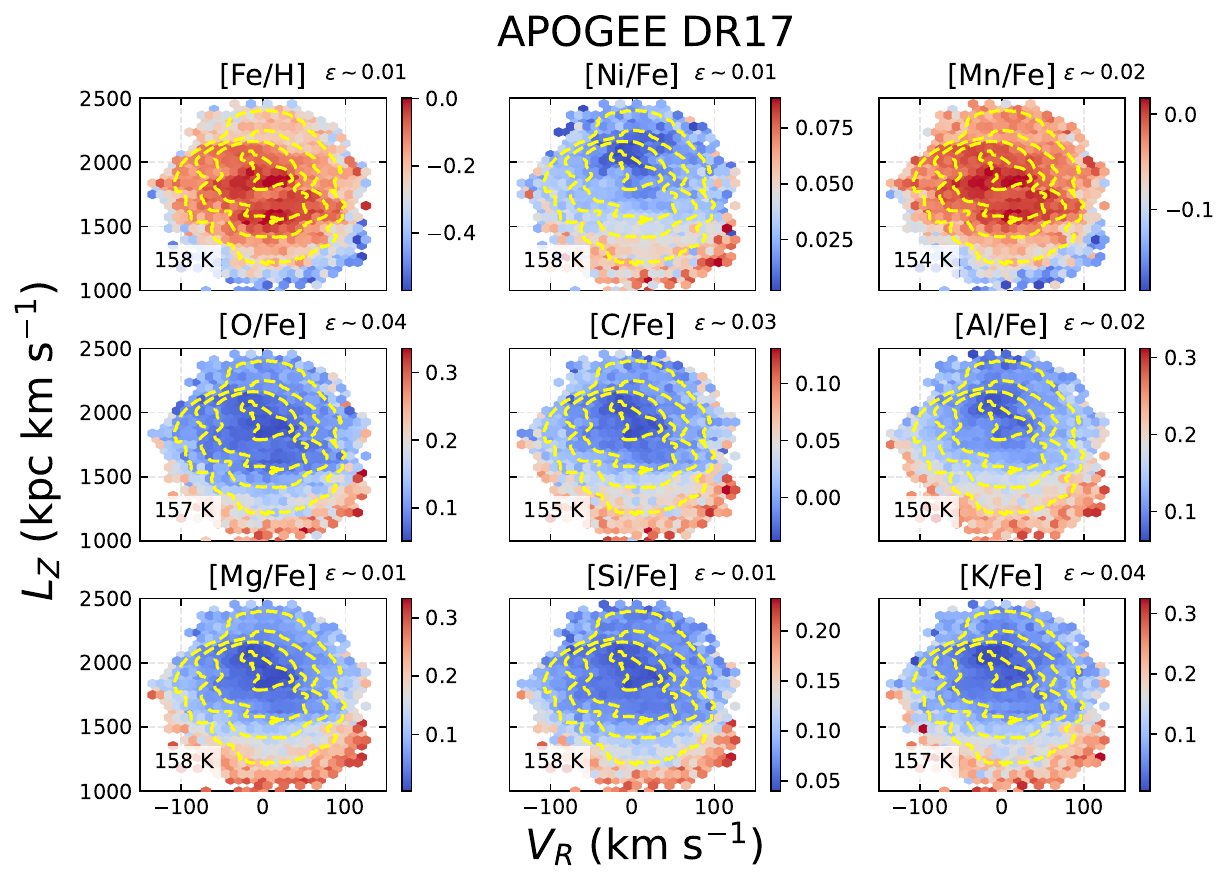}
    \caption{The $L_Z$-$V_R$ kinematics space coloured by 9 reliable element abundances in APOGEE DR17. Similar to Fig.\,\protect\ref{fig:lzvr_X_galah}, the flagged stellar abundances in each hexagonal bin are averaged and coloured, but to a different scale. Similar to GALAH, the chemical trend in Ni is different to that of Fe and Mn. APOGEE plots contain about half the number of stars in GALAH plots. The $\epsilon$ given next to the abundance subtitles represents the median uncertainties of chemical abundances reported the APOGEE survey.}
    \label{fig:lzvr_X_apogee}
\end{figure*}

To verify the chemical trends in Fig.\,\ref{fig:lzvr_X_galah}, we colour the kinematics space with nine reliable APOGEE abundances in Fig.\,\ref{fig:lzvr_X_apogee}. The nine elements are divided into four categories: Fe, Ni, and Mn as iron peak elements at the top row; light elements O and C in the second row; Mg and Si in the bottom row are alpha elements; and on the bottom right are Odd-Z elements Al and K. Neutron capture process elements are not included as their measurements are considered unreliable in APOGEE. 

We report that the APOGEE results show a good match with the GALAH results. The kinematics space coloured by common abundances, Fe, Ni, Mn, O, Al, Mg, and Si, are nearly identical between GALAH and APOGEE. However, the sharp transition into the low $L_Z$ patch in Ni and Al is not well recognised. The minimum at Sirius is as noticeable in Ni, and Al in APOGEE. Moreover, with higher precision, alpha elements Mg and Si now also show hints of an anti-peak at Sirius. The two new elements, C and K show patterns nearly identical to that of Al, with Sirius as the minima and enhancement in the background.

Overall, we observe that the Hercules I and II follow chemical trends similar to high $L_Z$ kinematic groups around LSR in both APOGEE and GALAH surveys. They are relatively metal-rich, alpha-poor, and s-process-rich young stars. In both surveys, the two most significant features identified are the Sirius group and the low $L_Z$ patch at Hercules III and IV. The Sirius group emerges as the global minimum in Odd-Z elements and age, the global maximum in s-process elements, and a local minimum in alpha elements. On the other hand, the low $L_Z$ patch is enhanced in Odd-Z elements and deficient in Ba and Li in GALAH, but not well observed in APOGEE. Note that Ni and V, while theoretically belonging to the iron peak \citep{Kobayashi2020ApJ...900..179K}, show global patterns over $L_Z$ opposite to Fe and Mn and local patterns more similar to Odd-Z elements, Na and Al.

\newcommand{\uu}{\color{red}$\Uparrow$\color{black}}
\newcommand{\dd}{\color{blue}$\Downarrow$\color{black}}
\newcommand{\mm}{$\Rightarrow$}

\begin{table}
    \centering
    \begin{tabular}{| c ||c|c|c|c|c|c|}
        \hline
         Group              & $\alpha$  & Fe    & Ni \& V   & Odd-Z & s-process & age   \\
         \hline
         Arch/Hat           & \dd       & \mm   & \dd       & \dd   & \uu       & \dd   \\
         Sirius             & \dd       & \uu   & \dd       & \dd   & \uu       & \dd   \\
         Hyades             & \dd       & \uu   & \dd       & \dd   & \uu       & \dd   \\
         Herc. I \& II   & \dd       & \uu   & \dd       & \dd   & \uu       & \dd   \\
         Herc. III \& IV & \mm       & \uu   & \uu       & \uu   & \dd       & \mm   \\
         Background         & \uu       & \dd   & \uu       & \uu   & \dd       & \uu   \\
         \hline
    \end{tabular}
    \caption{A summary of relative chemical and age trends among kinematic groups in Fig.\,\ref{fig:lzvr_X_galah}. Enhancements are marked by red up arrows \uu; deficiencies are marked by blue down arrows \dd; less prominent transition zones are marked by black right arrows \mm.}
    \label{tab:trends}
\end{table}

A brief summary of the chemical trends among kinematic groups in Fig.\,\ref{fig:lzvr_X_galah} is presented in Table \ref{tab:trends}. Noticeably, the sharp transition in abundances happening at $L_Z \sim 1600$\,kpc\,km\,s\inv results in a low $L_Z$ patch at Hercules III and IV, which emerges as a unique feature that is highly enhanced in metals and Odd-Zs. Combine with the [Mg/Fe]-[Fe/H] plots in Figures \ref{fig:mgfe_galah} and \ref{fig:mgfe_apogee}, we conclude that the low $L_Z$ background is mainly comprised of alpha-rich, Fe-poor, Odd-Z-rich, and s-deficient thick disc stars. On the other hand, the high $L_Z$ kinematic groups (Hat, Sirius, and Hyades) and the top Hercules subgroups I and II include majoritively alpha-poor, metal-rich, Odd-Z-poor, and s-enhanced thin disc stars. Thus, the alpha-rich, Fe-rich, Odd-Z enhanced, and relatively s-deficient low $L_Z$ patch acts as a peculiar group in the SNd. 

\begin{figure}
    \centering
    \includegraphics[width=\linewidth]{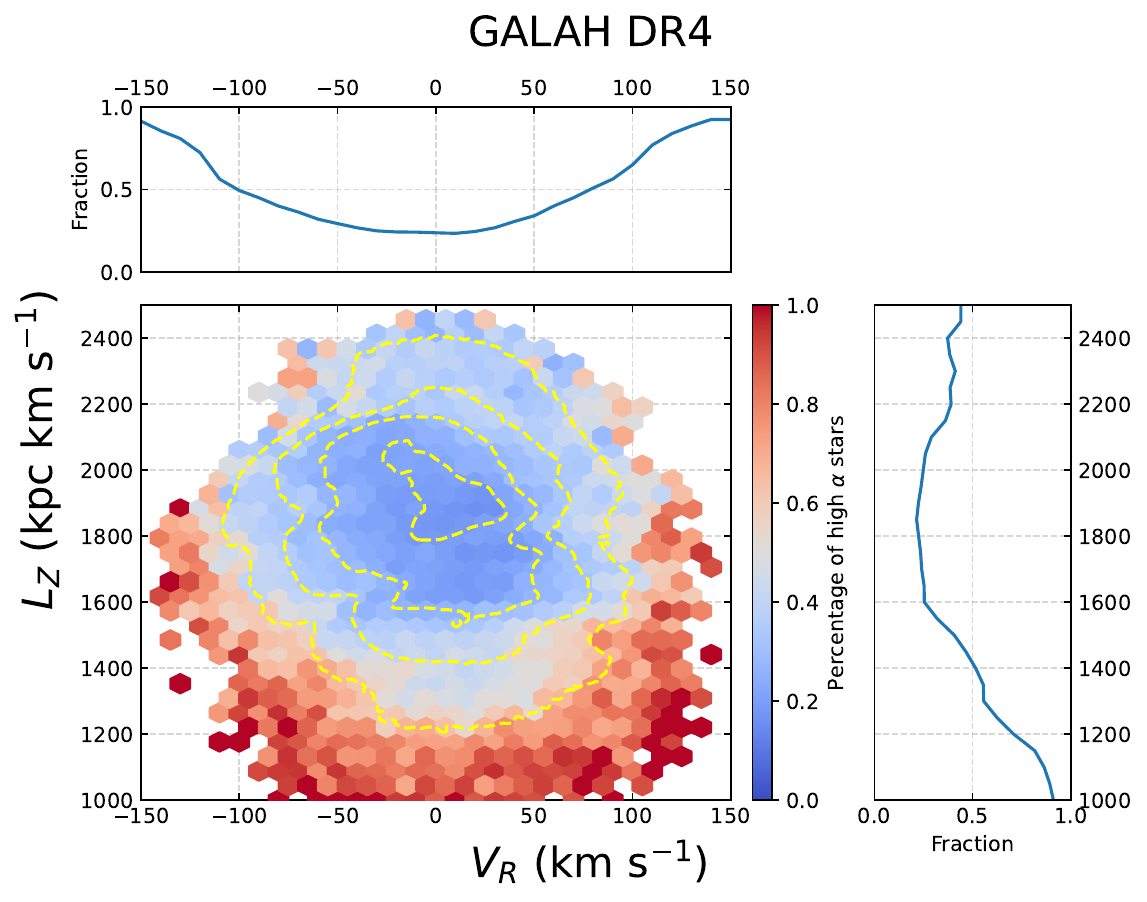}
    \caption{The fraction of high alpha stars ([Mg/Fe] $> 0.1$) in the $L_Z$-$V_R$ kinematic space in GALAH. The fraction of high alpha thick disc stars rises significantly from $25.4$\,per\,cent at $L_Z = 1600$\,kpc\,km\,s\inv to $70.9$\,per\,cent at $L_Z = 1200$\,kpc\,km\,s\inv, whereas the high $L_Z$ groups around the LSR and Hercules I and II are less contaminated, with a thick disc fraction of $23.6$\,per\,cent. The similarity of this coloured $L_Z$-$V_R$ space and many panels in Figs.\,\ref{fig:lzvr_X_galah} and \ref{fig:lzvr_X_apogee} reveal the significant effect of high alpha stars on the trends. The star density contours from Fig.\,\ref{fig:selection} are plotted as yellow dashed lines. The side plots shows the proportion of thick disc stars as a function of $L_Z$ and $V_R$ respectively.}
    \label{fig:Td_contamination}
\end{figure}

While Hercules III and IV are very Fe-rich, they show signatures different to the LSR and other kinematic groups in alpha elements, s-process elements and age. In almost all these elements, Hercules III and IV appear similar to background stars or as a transition zone between the main patch and background. Although this may indicate Hercules III and IV as a new population of disc stars, Figures \ref{fig:mgfe_galah} and \ref{fig:mgfe_apogee} recognise the potential effect of the contamination of thick disc stars in Hercules III and IV that may undermine the super high metal enrichment in the low $L_Z$ patch. In fact, the $L_Z$-$V_R$ kinematic space coloured by the percentage of high alpha thick disc stars ([Mg/Fe] $> 0.1$) in Fig.\,\ref{fig:Td_contamination} shows a worrisome similarity to V, Al, and alpha panels in Fig.\,\ref{fig:lzvr_X_galah} and \ref{fig:lzvr_X_apogee}. High alpha stars dominate at lower $L_Z$ and higher $|V_R|$. In the low $L_Z$ groups (Hercules III \& IV), the increased contamination of thick disc population and the significant difference between two populations can dominate the kinematic-abundance space and undermine any useful chemical features in the kinematic groups. Although not presented, the figure of APOGEE stars is almost identical to that of GALAH.

Moreover, as the low $L_Z$ patch is not well identified in APOGEE and there exist the inconsistency between iron-peak elements, the existence and properties of this feature need to be carefully considered.

Overall, the chemically colour-coded $L_Z$-$V_R$ kinematic space shows trends consistent among most kinematic groups (Hat, Sirius, Hyades, and Hercules I \& II). These groups are dominated by thin disc stars that are alpha-deficient and metal-enhanced. In addition, a low $L_Z$ patch at Hercules III \& IV shows a trend different to both the other kinematic groups and the background. However, the inconsistencies among Fe-peak elements and between GALAH and APOGEE surveys require a careful investigation into these conclusions.

\subsection{Variation of kinematic structures with abundances}\label{sec:x-cut}

\begin{figure*}
    \centering
    \includegraphics[width=\linewidth]{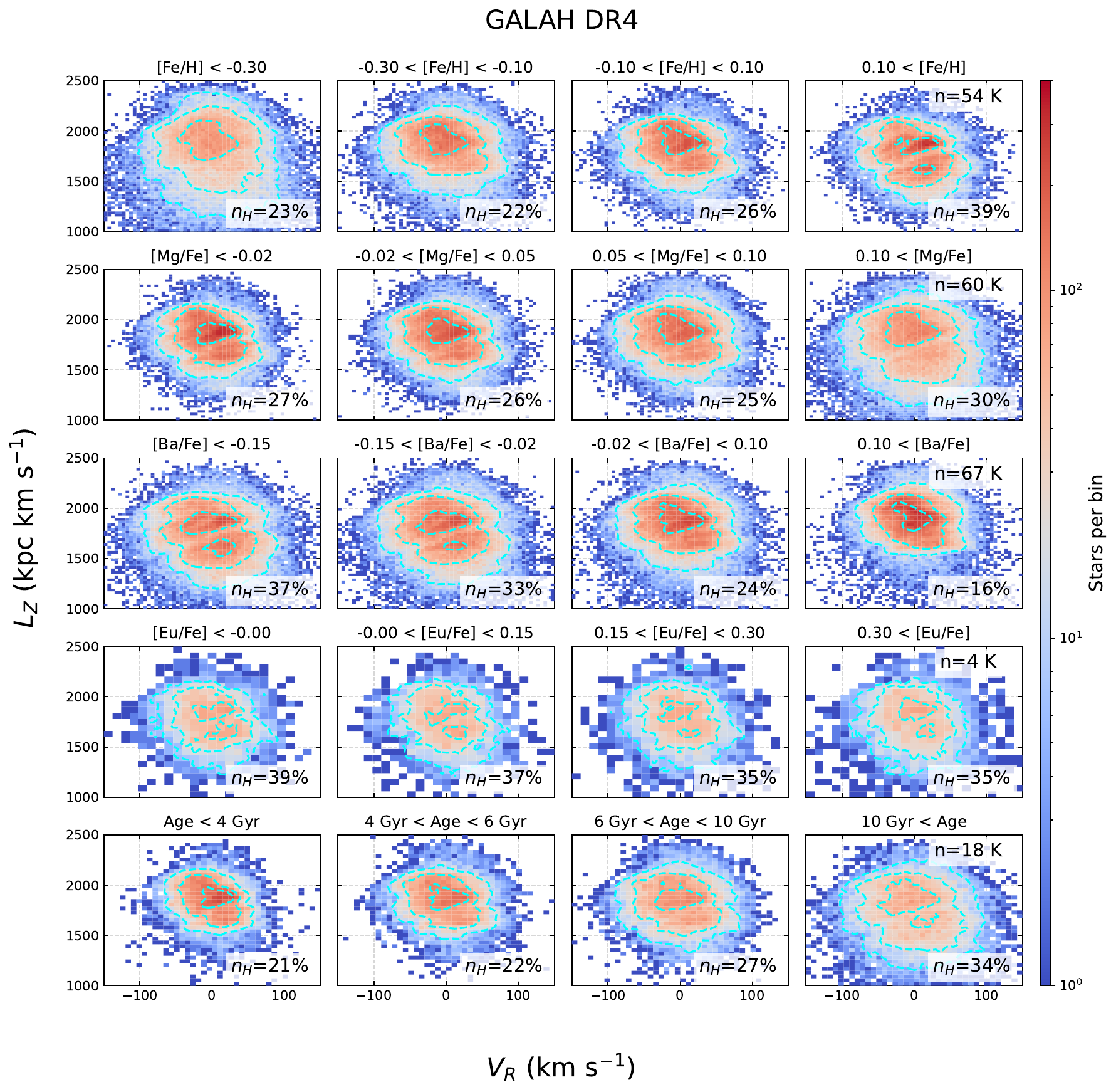}
    \caption{The 2D histogram of four intervals of four GALAH abundances and isochrone age in the SNd. The high alpha, low Fe thick disc population appears to be a near-flat distribution that contributes little to the kinematic structures. Top row: [Fe/H]; second row: [Mg/Fe]; third row: [Ba/Fe]; fourth row: [Eu/Fe]; bottom row: age. Lower abundances are on the left and higher are on the right. The number of stars for plots in each row is randomly truncated to be identical, marked at the top right on the rightmost plot of each row. The three dashed cyan contours encircle $90\text{\,per\,cent}$, $70\text{\,per\,cent}$, and $25\text{\,per\,cent}$ of stars in the plot. The portion of Hercules stars is marked on the bottom right.}
    \label{fig:lzvr_cut_galah}
\end{figure*}

To consider the effect of the thick disc population on the abundances of the low $L_Z$ patch at Hercules III and IV, a simple approach would be cutting off the stars with [Mg/Fe] higher than a selected threshold. However, although the randomness and higher velocity dispersion of thick disc stars suggest that they are less likely to contribute to stable kinematic structures, their contribution is uncertain and need careful consideration. Therefore, in this section, we attempt to understand the kinematic structure of stars with different abundances. We partition abundance spaces into bins and examine the distribution of stars in the $L_Z$-$V_R$ kinematics plane in each bin. We will show that the alpha-enhanced, Fe-deficient thick disc population is flat in the $L_Z$-$V_R$ kinematic space and hence has minor contribution to the kinematic structures in the SNd.

First, we examine the kinematic distribution in the chemical space of GALAH stars in the SNd, including information on iron-peak, alpha, s-process and, r-process elements in addition to stellar age. Fig.\,\ref{fig:lzvr_cut_galah} shows the $L_Z$-$V_R$ kinematics distribution of GALAH stars in abundance intervals. In each row, we partition the SNd stars into four bins in an abundance space. In each row, the sample size in the bins is normalised by random removal, similar to that in \S\,\ref{sec:k-group}. The portion of Hercules stars is marked on the bottom right of each bin. The dashed cyan contours encircle $90\text{\,per\,cent}$, $70\text{\,per\,cent}$, and $25\text{\,per\,cent}$ of stars in the respective bin. We choose four chemical and age spaces to present: [Fe/H] as the most typical iron-peak element; [Mg/Fe] as the most typical alpha; [Ba/Fe] as an s-process representative; [Eu/Fe] as an r-process representative; and stellar ages. 

\begin{figure*}
    \centering
    \includegraphics[width=\linewidth]{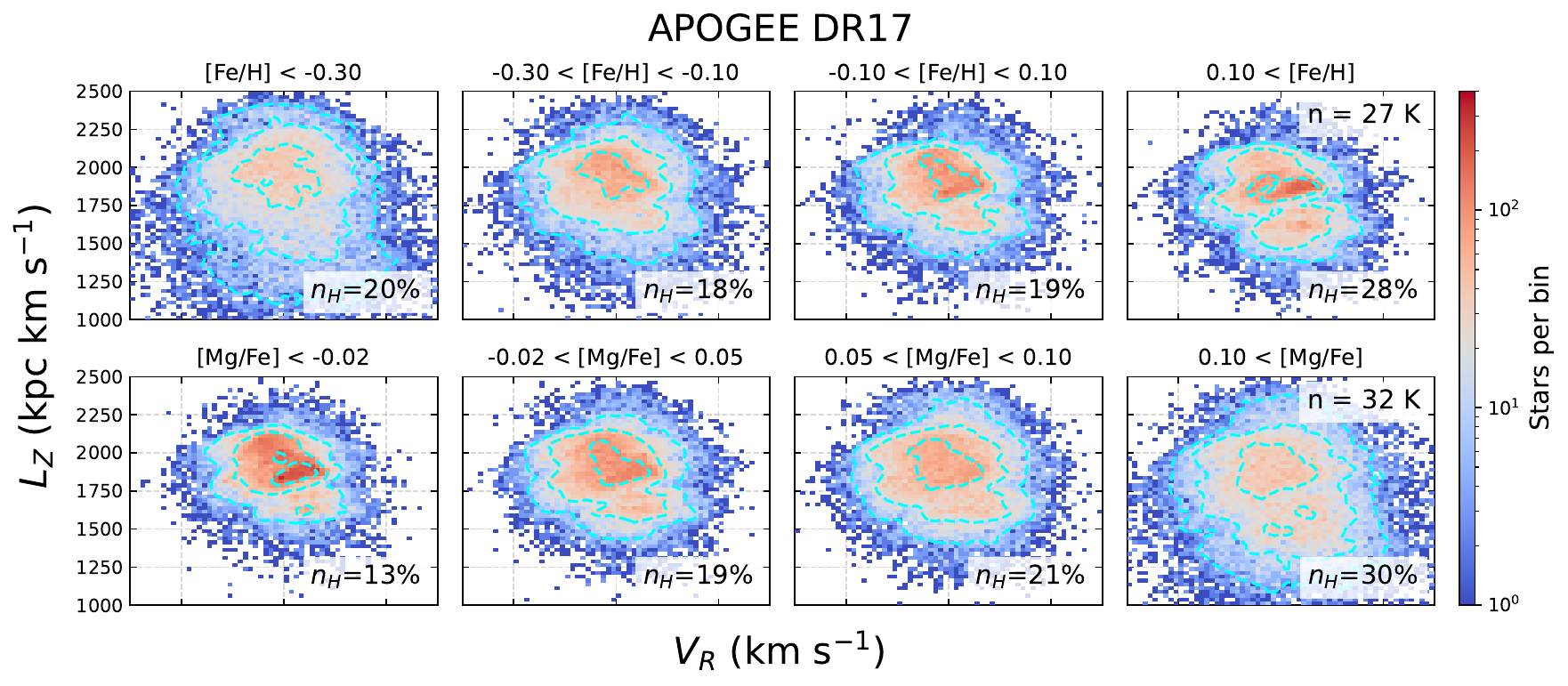}
    \caption{The kinematics of four intervals of two APOGEE abundances in the SNd. Similar to Fig.\,\ref{fig:lzvr_cut_galah}, the thick disc population shows little kinematic structure. Top row: [Fe/H]; Bottom row: [Mg/Fe]. Lower abundances are on the left and higher are on the right.}
    \label{fig:lzvr_cut_apogee}
\end{figure*}

\begin{itemize}
    \item \textbf{[Fe/H]}: At the top row, the SNd stars are partitioned into Fe bins. The lowest [Fe/H] bin, while showing a top and a bottom patch, shows almost no fine substructures identified in \S\,\ref{sec:kinematics} except a hat-like Sirius, a well-populated Hyades-LSR, and a hint of Hercules II. At intermediate [Fe/H], Hercules III and IV regions become underpopulated as thick disc stars decrease in portion, yet the total number of Hercules stars increases as kinematic groups like Hercules I, II, Hyades, and Sirius sharpen out. At the most metal-rich bin, Hercules III and IV appear again as the Horn stands out at the top. The portion of Hercules stars reached its maximum in the rightmost bin, in which Hercules occupies almost $40 \text{\,per\,cent}$ of stars in the most metal-rich context.

    \item \textbf{[Mg/Fe]}: In the second row, we show the Mg bins. In this row, the kinematic groups are most sharpened in the most alpha-poor bin. However, the significance of the Horn and Hercules III and IV remains largely constant throughout the bins. Similarly, the portion of Hercules and overdensities at Hercules I and II go through little evolution. The fraction of Hercules stars $n_H$ decreases slightly with the enrichment in Mg and then increases significantly due to the appearance of high alpha thick disc stars on the high end of Mg bins. When [Mg/Fe] $> 0.1$\,dex, the stars distribute almost uniformly, similar to that in the most Fe-poor bin.

    \item \textbf{[Ba/Fe]}: In the middle row where [Ba/Fe] bins are shown, $n_H$ decreases steadily from the Ba-poor left to the Ba-rich right. The constant sharpness of Hercules I and II throughout the bins attributes the decrease to the disappearance of the low $L_Z$ patch and the increase in the significance of the Sirius peak. We also notice the high concentration in the Horn at lower Ba bins. 

    \item \textbf{[Eu/Fe]}: As expected, the kinematic space shows no evidence of any variation in the distribution between different Eu bins, and the bins show almost no structures except a top and a bottom patch.

    \item \textbf{Age}: In the bottom row, the age bins show an evolution in Hercules portion and the population of the low $L_Z$ patch opposite to Ba: Hercules III and IV disappear in the youngest stars and populate gradually in the intermediate age bins, while the portion of Hercules increases with the appearing of Hercules III and IV. Similar to the highest Mg bin, the kinematic space comprised of the oldest stars is near-uniform as the high alpha thick disc stars.
\end{itemize}

We conduct a similar analysis on the APOGEE stars in Fig.\,\ref{fig:lzvr_cut_apogee}. We choose the same bins as in the GALAH analysis, except we limit our elements to iron and magnesium. The evolution of the kinematics structures in the abundance spaces in APOGEE agrees well with that of GALAH. 

\begin{itemize}
    \item \textbf{[Fe/H]}: We observe the near-uniform most iron-poor bin, a decrease in $n_H$ due to the absence of background stars at Hercules III and IV, and the sharpening of Hercules III, IV, and the Horn along with the increase in the total Hercules portion at the most metal-rich bin.
    
    \item \textbf{[Mg/Fe]}: Similarly in Mg bins, the near uniform high alpha stars well define the thick disc. However, the low $L_Z$ Hercules substructures show a strong trend along [Mg/Fe]. While Hyades, LSR, Sirius and high $L_Z$ Hercules structures show their sharpest state in the most alpha-poor bin, the low $L_Z$ patch is nearly completely absent. In APOGEE, the arrival of Hercules III and IV with the increase of [Mg/Fe] forms the steady increase in $n_H$ with Mg enrichment.
\end{itemize}

In general, the examination of the abundances and age spaces of APOGEE and GALAH stars make qualitative agreements on the inhabitation of the Hercules group as a whole and the variations of substructures. As a whole, the fraction of the Hercules group increases with SNe Ia element Fe, decreases with AGB element Ba, increases with age, and is independent of Neutron star merging element Eu. A disagreement is found in the variation of SNe type II element Mg: $n_H$ increases in APOGEE while remaining nearly constant in GALAH. For substructures, Hercules I and II are observed in nearly all bins except in high Mg, low Fe, and high stellar age thick disc bins in which the structures are smoothened. Hercules III and IV show significantly different behaviours. They start uniform like the other two subgroups in the metal-poor end, but decrease in the intermediate iron abundances, and emerge at the very metal-rich end [Fe/H] $> 0.1$ dex. The low $L_Z$ structure well-populates the Ba-poor end, decreases significantly, and almost vanishes in the Ba-rich bin. To resolve the conflict in the Mg variations between GALAH and APOGEE, we favour the result in APOGEE slightly more due to the better distinguishability between the high and low alpha populations in \S\,\ref{sec:k-group}.

\subsection{The alpha-deficient sample}
\label{sec:x-k-td}

\begin{figure*}
    \centering
    \includegraphics[width=\linewidth]{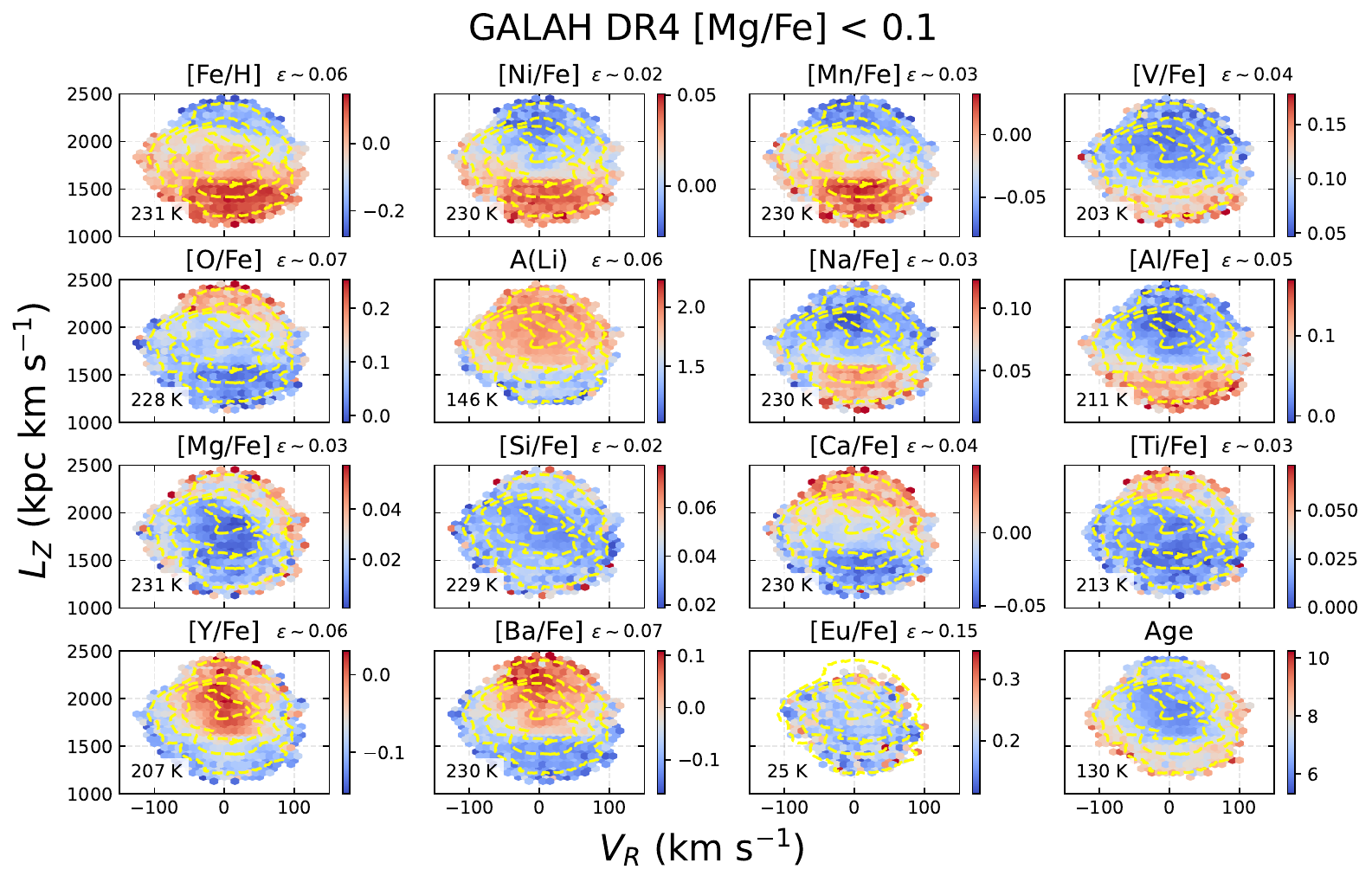}
    \caption{The $L_Z$-$V_R$ kinematics space coloured by 15 reliable chemical abundances and age in GALAH DR4 with [Mg/Fe] $<0.1$. Same to Fig.\,\protect\ref{fig:lzvr_X_galah} except with the removal of high alpha thick disc stars. The inconsistency in Fe-peak elements is resolved. Hercules III and IV stand out as the most Fe-peak-rich and Odd-Z-rich structure in the SNd.}
    \label{fig:x_k_lowmg_galah}
\end{figure*}

Regardless of minor disagreements between the two surveys, we can verify that the high alpha old sequence is nearly uniform in the kinematics space, and hence they contribute little to the kinematics structures we are interested in. Therefore, we choose a threshold at [Mg/Fe] $=0.1$\,dex based on the analysis in \S\,\ref{sec:x-cut} and discard all stars with [Mg/Fe] $\geq 0.1$\,dex. While this simple method cannot remove all thick disc stars and certainly removes some stars that have a thin disc origin, it is sufficient to reveal the true chemical features of the low-alpha low $L_Z$ patch at Hercules III and IV: a super Fe-peak-enhanced, Odd-Z enhanced, Alpha-deficient thin disc sample that is slightly older than other thin disc stars. These stars are likely the youngest stars in the old sequence and originated from the inner Galaxy.

Fig.\,\ref{fig:x_k_lowmg_galah} shows the same abundance-coloured $L_Z$-$V_R$ kinematics space as in Fig.\,\ref{fig:lzvr_X_galah}, but the high alpha stars are removed and the colour bars are modified. Similarly, the trends in Fig.\,\ref{fig:x_k_lowmg_galah} are summarised in Table \ref{tab:trends_td}. 

With the removal of high alpha stars, the discrepancy in iron-peak elements disappears. In this low alpha sample, the general trends of Fe and Mn are flipped, and all iron-peak abundances show a decreasing pattern over $L_Z$. This reflects the negative Galactic radial metallicity gradient: as $L_Z$ increases with guiding radii $R_g$, higher $L_Z$ indicates an orbit with larger $R_g$, and hence originate from a larger Galactic radius. Among alpha elements in the third row, the patterns in Mg and Si become less distinguishable, which can be expected as the alpha abundance spaces are limited by the removal of stars enhanced in the typical alpha element Mg. The other two alphas, Ca and Ti, show an inversion in the general trend in a similar way to Fe and Mn. As a light element close to alphas, the global pattern of O is also inverted. The abundance in O, Li, Ca, and Ti now all increase with $L_Z$. Patterns in Odd-Z elements and s-process elements sharpened slightly but do not show major variations in the global trends. No observable changes are found in the relative pattern of Eu and stellar ages.

In addition to the uniformity of behaviours in the iron-peak elements, another significant pattern observed is the emergence of the sharp transition at the low $L_Z$ patch in more abundances. In this low alpha set, the low $L_Z$ patch is observed to be the maxima in Fe, Ni, and Mn. The evidence is slightly weaker in V. The transition at the low $L_Z$ patch sharpens in Al, in agreement with the other Odd-Z element Na. Moreover, the low $L_Z$ patch becomes observable as the minima in O, Li, Ca, vaguely Ti, and Ba.

\begin{figure*}
    \centering
    \includegraphics[width=.75\linewidth]{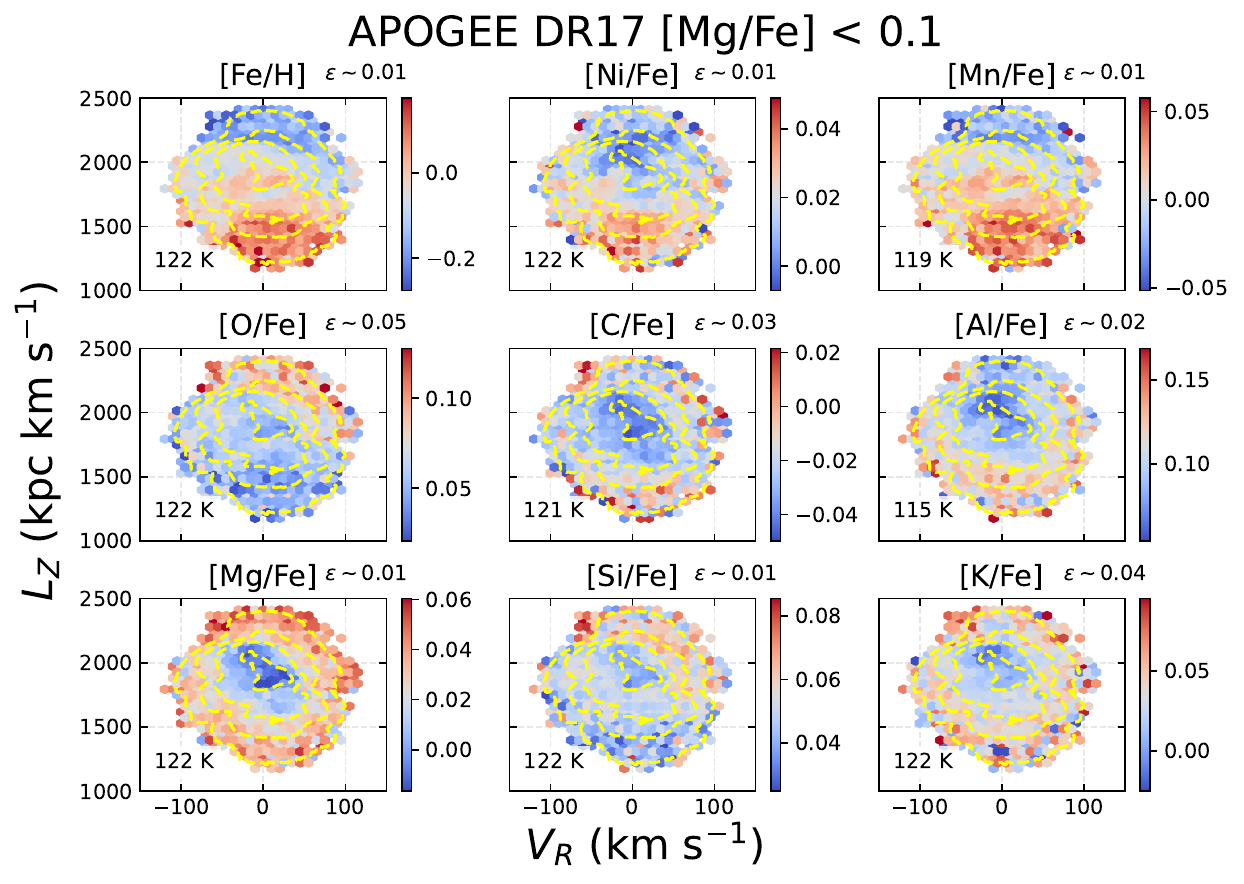}
    \caption{The $L_Z$-$V_R$ kinematics space coloured by 9 reliable element abundances in APOGEE DR17 stars with [Mg/Fe] $<0.1$. Same to Fig.\,\protect\ref{fig:lzvr_X_apogee} except with the removal of high alpha thick disc stars. Like Fig.\,\ref{fig:x_k_lowmg_galah}, the inconsistency in Fe-peak elements is resolved and Hercules III and IV appear as a spectial Fe-peak-rich low $L_Z$ patch.}
    \label{fig:x_k_lowmg_apogee}
\end{figure*}

\begin{table}
    \centering
    \begin{tabular}{| c ||c|c|c|c|c|c|}
        \hline
         Group              & $\alpha$  & Fe    & Ni \& V   & Odd-Z & S-process & age   \\
         \hline
         Arch/Hat           & \uu       & \dd   & \dd       & \mm   & \uu       & \dd   \\
         Sirius             & \dd       & \mm   & \dd       & \dd   & \uu       & \dd   \\
         Hyades             & \dd       & \uu   & \mm       & \dd   & \uu       & \dd   \\
         Herc. I \& II   & \dd       & \uu   & \mm       & \mm   & \uu       & \dd   \\
         Herc. III \& IV & \dd       & \uu\uu& \uu       & \uu   & \dd       & \uu   \\
         \hline
    \end{tabular}
    \caption{A summary of relative chemical and age trends among kinematic groups in Fig.\,\ref{fig:x_k_lowmg_galah}. Same as Table \ref{tab:trends} except with high Mg stars removed. The background group vanishes due to the removal. The \uu\uu{} emphasises the super Fe-enhanced low $L_Z$ patch at Hercules III and IV. Check Tab.\,\ref{tab:galah_stat} and \ref{tab:apogee_stat} for a more quantitative statistical presentation.}
    \label{tab:trends_td}
\end{table}


\newcommand{\rr}[1]{\color{red} #1 \color{black}}
\newcommand{\bb}[1]{\color{blue} #1 \color{black}}
\newcommand{\kk}[1]{#1}

\begin{table*}
    \renewcommand{\arraystretch}{1.5} 
    \setlength{\tabcolsep}{3pt} 
    
    \centering
    \begin{tabular}{cccccccccccc}
        \hline
        Abundance & Arch/Hat & Sirius & LSR & Hyades & Hercules I & Hercules II & Hercules III & Hercules IV & Background & $\mu \pm \sigma$ \\
        \hline
        \textbf{Iron-peak}\\

        [Fe/H] & $\bb{-0.18}\,^{-0.03}_{-0.32}$ & $\kk{-0.04}\,^{+0.06}_{-0.15}$ & $\kk{-0.01}\,^{+0.09}_{-0.12}$ & $\rr{+0.04}\,^{+0.15}_{-0.08}$ & $\rr{+0.03}\,^{+0.12}_{-0.08}$ & $\rr{+0.04}\,^{+0.16}_{-0.07}$ & $\rr{+0.12}\,^{+0.23}_{+0.01}$ & $\rr{+0.12}\,^{+0.23}_{+0.01}$ & $\bb{-0.39}\,^{-0.15}_{-0.59}$ & $-0.06\pm0.23$ \\
        
        [Ni/Fe] & $\bb{-0.01}\,^{+0.02}_{-0.05}$ & $\bb{-0.02}\,^{+0.02}_{-0.05}$ & $\bb{-0.01}\,^{+0.03}_{-0.05}$ & $\kk{+0.00}\,^{+0.04}_{-0.03}$ & $\kk{+0.01}\,^{+0.04}_{-0.02}$ & $\kk{+0.02}\,^{+0.05}_{-0.01}$ & $\rr{+0.04}\,^{+0.07}_{+0.01}$ & $\rr{+0.04}\,^{+0.07}_{+0.01}$ & $\rr{+0.04}\,^{+0.06}_{+0.01}$ & $+0.01\pm0.05$ \\
        
        [Mn/Fe] & $\bb{-0.07}\,^{-0.03}_{-0.10}$ & $\kk{-0.04}\,^{+0.01}_{-0.08}$ & $\kk{-0.03}\,^{+0.02}_{-0.07}$ & $\rr{-0.01}\,^{+0.05}_{-0.06}$ & $\kk{-0.02}\,^{+0.03}_{-0.06}$ & $\rr{-0.01}\,^{+0.05}_{-0.06}$ & $\rr{+0.02}\,^{+0.08}_{-0.04}$ & $\rr{+0.01}\,^{+0.07}_{-0.04}$ & $\bb{-0.13}\,^{-0.07}_{-0.18}$ & $-0.04\pm0.08$ \\
        
        [V/Fe] & $\kk{+0.06}\,^{+0.13}_{+0.01}$ & $\bb{+0.04}\,^{+0.12}_{-0.01}$ & $\kk{+0.05}\,^{+0.12}_{-0.01}$ & $\kk{+0.05}\,^{+0.13}_{-0.01}$ & $\kk{+0.05}\,^{+0.14}_{+0.00}$ & $\kk{+0.06}\,^{+0.15}_{+0.01}$ & $\kk{+0.10}\,^{+0.20}_{+0.04}$ & $\kk{+0.10}\,^{+0.21}_{+0.04}$ & $\rr{+0.19}\,^{+0.27}_{+0.12}$ & $+0.08\pm0.12$ \\

        \hline
        \textbf{Light}\\

        [O/Fe] & $\kk{+0.18}\,^{+0.28}_{+0.08}$ & $\kk{+0.12}\,^{+0.21}_{+0.00}$ & $\kk{+0.10}\,^{+0.20}_{-0.01}$ & $\kk{+0.08}\,^{+0.18}_{-0.04}$ & $\kk{+0.09}\,^{+0.19}_{-0.01}$ & $\kk{+0.08}\,^{+0.18}_{-0.03}$ & $\bb{+0.03}\,^{+0.14}_{-0.08}$ & $\bb{+0.04}\,^{+0.15}_{-0.07}$ & $\rr{+0.36}\,^{+0.57}_{+0.15}$ & $+0.12\pm0.24$ \\
        
        A(Li) & $\kk{+2.12}\,^{+2.32}_{+1.68}$ & $\kk{+2.15}\,^{+2.38}_{+1.68}$ & $\kk{+2.18}\,^{+2.41}_{+1.69}$ & $\kk{+2.13}\,^{+2.38}_{+1.60}$ & $\kk{+2.07}\,^{+2.33}_{+1.51}$ & $\kk{+2.02}\,^{+2.31}_{+1.43}$ & $\kk{+1.73}\,^{+2.18}_{+1.07}$ & $\bb{+1.62}\,^{+2.11}_{+0.98}$ & $\bb{+1.24}\,^{+1.63}_{+0.80}$ & $+1.96\pm0.72$ \\
        
        \hline
        \textbf{Odd-Z}\\

        [Na/Fe] & $\kk{+0.03}\,^{+0.08}_{-0.01}$ & $\kk{+0.01}\,^{+0.07}_{-0.05}$ & $\kk{+0.02}\,^{+0.08}_{-0.04}$ & $\kk{+0.03}\,^{+0.10}_{-0.03}$ & $\kk{+0.04}\,^{+0.11}_{-0.01}$ & $\kk{+0.06}\,^{+0.12}_{-0.00}$ & $\rr{+0.09}\,^{+0.15}_{+0.02}$ & $\rr{+0.08}\,^{+0.15}_{+0.02}$ & $\rr{+0.07}\,^{+0.12}_{+0.02}$ & $+0.03\pm0.10$ \\
        
        [Al/Fe] & $\kk{+0.03}\,^{+0.11}_{-0.05}$ & $\bb{-0.00}\,^{+0.08}_{-0.08}$ & $\bb{+0.01}\,^{+0.09}_{-0.07}$ & $\kk{+0.02}\,^{+0.10}_{-0.06}$ & $\kk{+0.04}\,^{+0.12}_{-0.05}$ & $\kk{+0.06}\,^{+0.13}_{-0.02}$ & $\rr{+0.12}\,^{+0.19}_{+0.04}$ & $\rr{+0.13}\,^{+0.19}_{+0.05}$ & $\rr{+0.26}\,^{+0.37}_{+0.15}$ & $+0.07\pm0.15$ \\

        \hline
        \textbf{Alpha}\\

        [Mg/Fe] & $\kk{+0.04}\,^{+0.07}_{+0.01}$ & $\kk{+0.02}\,^{+0.05}_{-0.02}$ & $\kk{+0.01}\,^{+0.05}_{-0.03}$ & $\bb{+0.01}\,^{+0.04}_{-0.03}$ & $\kk{+0.01}\,^{+0.05}_{-0.03}$ & $\kk{+0.01}\,^{+0.05}_{-0.03}$ & $\kk{+0.02}\,^{+0.05}_{-0.02}$ & $\kk{+0.02}\,^{+0.06}_{-0.02}$ & $\rr{+0.26}\,^{+0.35}_{+0.14}$ & $+0.04\pm0.11$ \\
        
        [Si/Fe] & $\kk{+0.04}\,^{+0.07}_{+0.01}$ & $\kk{+0.03}\,^{+0.05}_{-0.00}$ & $\kk{+0.03}\,^{+0.05}_{+0.00}$ & $\kk{+0.03}\,^{+0.05}_{+0.00}$ & $\kk{+0.03}\,^{+0.05}_{+0.00}$ & $\kk{+0.03}\,^{+0.06}_{+0.00}$ & $\kk{+0.03}\,^{+0.06}_{+0.01}$ & $\kk{+0.03}\,^{+0.06}_{+0.01}$ & $\rr{+0.19}\,^{+0.26}_{+0.10}$ & $+0.05\pm0.08$ \\
        
        [Ca/Fe] & $\kk{+0.02}\,^{+0.06}_{-0.02}$ & $\kk{+0.00}\,^{+0.05}_{-0.04}$ & $\kk{-0.00}\,^{+0.04}_{-0.05}$ & $\kk{-0.01}\,^{+0.03}_{-0.06}$ & $\kk{-0.01}\,^{+0.03}_{-0.06}$ & $\bb{-0.02}\,^{+0.02}_{-0.07}$ & $\bb{-0.04}\,^{+0.01}_{-0.09}$ & $\bb{-0.04}\,^{+0.01}_{-0.08}$ & $\rr{+0.14}\,^{+0.20}_{+0.05}$ & $+0.01\pm0.09$ \\
        
        [Ti/Fe] & $\kk{+0.04}\,^{+0.08}_{+0.01}$ & $\kk{+0.01}\,^{+0.05}_{-0.02}$ & $\kk{+0.00}\,^{+0.04}_{-0.03}$ & $\kk{-0.00}\,^{+0.04}_{-0.03}$ & $\kk{+0.00}\,^{+0.04}_{-0.02}$ & $\kk{+0.00}\,^{+0.04}_{-0.03}$ & $\kk{+0.00}\,^{+0.04}_{-0.03}$ & $\kk{+0.01}\,^{+0.04}_{-0.02}$ & $\rr{+0.22}\,^{+0.28}_{+0.11}$ & $+0.03\pm0.09$ \\

        \hline
        \textbf{S-process}\\

        [Y/Fe] & $\kk{-0.05}\,^{+0.03}_{-0.11}$ & $\kk{+0.00}\,^{+0.08}_{-0.07}$ & $\kk{+0.00}\,^{+0.09}_{-0.07}$ & $\kk{-0.01}\,^{+0.07}_{-0.08}$ & $\kk{-0.03}\,^{+0.04}_{-0.10}$ & $\kk{-0.05}\,^{+0.02}_{-0.11}$ & $\kk{-0.09}\,^{-0.03}_{-0.14}$ & $\bb{-0.10}\,^{-0.04}_{-0.15}$ & $\kk{-0.09}\,^{-0.01}_{-0.15}$ & $-0.04\pm0.14$ \\
        
        [Ba/Fe] & $\rr{+0.07}\,^{+0.17}_{-0.05}$ & $\rr{+0.08}\,^{+0.19}_{-0.05}$ & $\rr{+0.06}\,^{+0.17}_{-0.08}$ & $\kk{+0.01}\,^{+0.13}_{-0.11}$ & $\kk{-0.00}\,^{+0.11}_{-0.12}$ & $\kk{-0.04}\,^{+0.08}_{-0.15}$ & $\bb{-0.12}\,^{-0.03}_{-0.20}$ & $\bb{-0.12}\,^{-0.04}_{-0.19}$ & $\bb{-0.09}\,^{+0.01}_{-0.18}$ & $-0.02\pm0.18$ \\

        \hline
        \textbf{R-process}\\

        [Eu/Fe] & $\kk{+0.20}\,^{+0.36}_{+0.06}$ & $\kk{+0.20}\,^{+0.35}_{+0.05}$ & $\kk{+0.19}\,^{+0.34}_{+0.04}$ & $\kk{+0.17}\,^{+0.33}_{+0.04}$ & $\kk{+0.16}\,^{+0.32}_{+0.03}$ & $\kk{+0.16}\,^{+0.31}_{+0.03}$ & $\kk{+0.15}\,^{+0.32}_{+0.03}$ & $\kk{+0.16}\,^{+0.34}_{+0.03}$ & $\rr{+0.30}\,^{+0.44}_{+0.13}$ & $+0.19\pm0.23$ \\
        
        Age & $\kk{+6.75}\,^{+8.60}_{+5.39}$ & $\bb{+5.76}\,^{+7.69}_{+4.44}$ & $\bb{+5.62}\,^{+7.58}_{+4.29}$ & $\bb{+5.55}\,^{+7.42}_{+4.19}$ & $\kk{+6.10}\,^{+8.08}_{+4.60}$ & $\kk{+6.38}\,^{+8.33}_{+4.84}$ & $\kk{+7.49}\,^{+9.34}_{+5.75}$ & $\rr{+8.03}\,^{+9.65}_{+6.10}$ & $\rr{+12.59}\,^{+13.59}_{+10.45}$ & $+6.89\pm2.97$ \\

        \hline
    \end{tabular}
    \caption{Abundance statistics of GALAH stars with [Mg/Fe] $< 0.1$ (except the background sample that includes all stars). The 25 per cent and 75 per cent quantiles of the samples are marked on the top and bottom right of the median ($\mu_g$) as an uncertainty measure. Kinematic group samples with median values $\mu_g$ lie outside the range of the median $\mu \pm $ a third of the standard deviation $\frac{1}{3} \sigma$ of the entire SNd sample (shown in the last column) are highlighted with colours, blue as deficient and red as enhanced. For example, the whole SNd sample has median [Fe/H] $\mu = -0.06$ and standard deviation $\sigma = 0.23$. The median of the thin disc Arch/Hat group has median $\mu_{\text{arch/hat}} = -0.18 < -0.06-0.23/3 = -0.14$, so it is marked blue. On the other hand, as the median of the thin disc Hercules IV group $\mu_{\text{HerIV}} = 0.12 > -0.06 + 0.23/3 = 0.02$, it is marked red. The low $L_Z$ patch at Hercules III and IV is highlighted as a Fe-peak-rich, Odd-Z-rich, alpha-poor and lightly s-process-poor peculiar group.}
    \label{tab:galah_stat}
\end{table*}

\begin{table*}
    \renewcommand{\arraystretch}{1.5} 
    \setlength{\tabcolsep}{3pt} 

    \centering
    \begin{tabular}{cccccccccccc}
        \hline
        Abundance & Arch/Hat & Sirius & LSR & Hyades & Hercules I & Hercules II & Hercules III & Hercules IV & Background & $\mu \pm \sigma$ \\
        \hline
        \textbf{Iron-peak}\\

        [Fe/H] & $\bb{-0.18}\,^{-0.05}_{-0.31}$ & $\kk{-0.05}\,^{+0.04}_{-0.15}$ & $\kk{-0.01}\,^{+0.09}_{-0.13}$ & $\rr{+0.04}\,^{+0.14}_{-0.07}$ & $\kk{+0.01}\,^{+0.11}_{-0.12}$ & $\rr{+0.02}\,^{+0.14}_{-0.11}$ & $\rr{+0.09}\,^{+0.21}_{-0.05}$ & $\rr{+0.10}\,^{+0.21}_{-0.03}$ & $\bb{-0.44}\,^{-0.23}_{-0.62}$ & $-0.07\pm0.23$ \\
        
        [Ni/Fe] & $\kk{+0.02}\,^{+0.04}_{-0.01}$ & $\bb{-0.00}\,^{+0.03}_{-0.03}$ & $\kk{+0.01}\,^{+0.04}_{-0.02}$ & $\kk{+0.02}\,^{+0.05}_{-0.01}$ & $\kk{+0.02}\,^{+0.05}_{-0.00}$ & $\kk{+0.03}\,^{+0.05}_{+0.00}$ & $\kk{+0.03}\,^{+0.06}_{+0.01}$ & $\kk{+0.03}\,^{+0.05}_{+0.01}$ & $\rr{+0.06}\,^{+0.09}_{+0.02}$ & $+0.02\pm0.05$ \\
        
        [Mn/Fe] & $\kk{-0.02}\,^{+0.02}_{-0.06}$ & $\kk{-0.00}\,^{+0.04}_{-0.04}$ & $\kk{+0.01}\,^{+0.05}_{-0.03}$ & $\rr{+0.03}\,^{+0.07}_{-0.01}$ & $\kk{+0.02}\,^{+0.06}_{-0.02}$ & $\rr{+0.03}\,^{+0.06}_{-0.01}$ & $\rr{+0.04}\,^{+0.08}_{-0.00}$ & $\rr{+0.03}\,^{+0.07}_{-0.01}$ & $\bb{-0.12}\,^{-0.06}_{-0.18}$ & $-0.01\pm0.09$ \\

        \hline
        \textbf{Light}\\

        [O/Fe] & $\kk{+0.07}\,^{+0.14}_{+0.03}$ & $\kk{+0.05}\,^{+0.11}_{-0.01}$ & $\kk{+0.04}\,^{+0.10}_{-0.01}$ & $\kk{+0.04}\,^{+0.09}_{-0.01}$ & $\kk{+0.05}\,^{+0.10}_{+0.01}$ & $\kk{+0.05}\,^{+0.10}_{+0.00}$ & $\kk{+0.04}\,^{+0.08}_{+0.00}$ & $\kk{+0.05}\,^{+0.08}_{+0.01}$ & $\rr{+0.25}\,^{+0.35}_{+0.14}$ & $+0.07\pm0.11$ \\
        
        [C/Fe] & $\kk{-0.02}\,^{+0.03}_{-0.07}$ & $\kk{-0.04}\,^{+0.01}_{-0.09}$ & $\kk{-0.03}\,^{+0.01}_{-0.08}$ & $\kk{-0.03}\,^{+0.00}_{-0.08}$ & $\kk{-0.03}\,^{+0.01}_{-0.07}$ & $\kk{-0.02}\,^{+0.01}_{-0.07}$ & $\kk{-0.01}\,^{+0.02}_{-0.05}$ & $\kk{-0.01}\,^{+0.03}_{-0.04}$ & $\rr{+0.08}\,^{+0.13}_{+0.02}$ & $-0.01\pm0.08$ \\

        \hline
        \textbf{Odd-Z}\\

        [Al/Fe] & $\kk{+0.09}\,^{+0.14}_{+0.04}$ & $\bb{+0.05}\,^{+0.12}_{+0.01}$ & $\kk{+0.07}\,^{+0.13}_{+0.02}$ & $\kk{+0.07}\,^{+0.14}_{+0.02}$ & $\kk{+0.10}\,^{+0.16}_{+0.05}$ & $\kk{+0.10}\,^{+0.16}_{+0.05}$ & $\kk{+0.12}\,^{+0.18}_{+0.05}$ & $\kk{+0.12}\,^{+0.18}_{+0.04}$ & $\rr{+0.24}\,^{+0.29}_{+0.17}$ & $+0.10\pm0.10$ \\
        
        [K/Fe] & $\kk{+0.05}\,^{+0.10}_{-0.01}$ & $\kk{+0.01}\,^{+0.06}_{-0.05}$ & $\kk{+0.01}\,^{+0.07}_{-0.04}$ & $\kk{+0.01}\,^{+0.06}_{-0.04}$ & $\kk{+0.03}\,^{+0.08}_{-0.02}$ & $\kk{+0.03}\,^{+0.09}_{-0.02}$ & $\kk{+0.04}\,^{+0.09}_{-0.01}$ & $\kk{+0.05}\,^{+0.11}_{-0.01}$ & $\rr{+0.25}\,^{+0.32}_{+0.15}$ & $+0.05\pm0.15$ \\
        
        \hline
        \textbf{Alpha}\\

        [Mg/Fe] & $\kk{+0.05}\,^{+0.07}_{+0.01}$ & $\bb{-0.00}\,^{+0.04}_{-0.04}$ & $\bb{-0.00}\,^{+0.04}_{-0.04}$ & $\bb{-0.01}\,^{+0.03}_{-0.05}$ & $\kk{+0.02}\,^{+0.05}_{-0.02}$ & $\kk{+0.02}\,^{+0.06}_{-0.02}$ & $\kk{+0.04}\,^{+0.06}_{+0.00}$ & $\kk{+0.04}\,^{+0.07}_{+0.01}$ & $\rr{+0.28}\,^{+0.33}_{+0.19}$ & $+0.04\pm0.10$ \\
        
        [Si/Fe] & $\kk{+0.06}\,^{+0.09}_{+0.02}$ & $\kk{+0.04}\,^{+0.08}_{+0.01}$ & $\kk{+0.04}\,^{+0.08}_{+0.01}$ & $\kk{+0.04}\,^{+0.08}_{+0.01}$ & $\kk{+0.06}\,^{+0.09}_{+0.02}$ & $\kk{+0.05}\,^{+0.09}_{+0.02}$ & $\kk{+0.05}\,^{+0.08}_{+0.01}$ & $\kk{+0.04}\,^{+0.08}_{+0.01}$ & $\rr{+0.19}\,^{+0.23}_{+0.12}$ & $+0.06\pm0.07$ \\

        \hline
    \end{tabular}
    \caption{Abundance statistics of APOGEE stars. All the samples are restricted to stars with [Mg/Fe] $< 0.1$ execept for the background sample. Similar to Tab.\,\ref{tab:galah_stat}, kinematic group samples with median values $\mu_g$ lie outside the range of $\mu \pm \frac{1}{3} \sigma$ of the entire SNd sample are highlighted with colours. Most of the conclusions agree with that of Tab.\,\ref{tab:galah_stat}.}
    \label{tab:apogee_stat}
\end{table*}


The same removal is conducted on the APOGEE data and is presented in Fig.\,\ref{fig:x_k_lowmg_apogee}. The general trend in the iron-peak class synchronises, in the same way the discrepancies in GALAH abundances resolve. In addition to that, excitingly, the sharp transition at $L_Z \sim 1600$\,kpc\,km\,s\inv, which is indistinguishable in all nine abundances in Fig.\,\ref{fig:lzvr_X_apogee}, is well identified in the low alpha sample of APOGEE stars. In Fig.\,\ref{fig:x_k_lowmg_apogee}, the evidence of the low $L_Z$ patch can be observed as the global maxima in all iron-peak elements, Fe, Ni, and Mn. Although there is no apparent evidence found in Odd-Z and alpha elements, this resolves the major discrepancy between the two surveys on the existence of this low $L_Z$ patch at Hercules III and IV.

Therefore, we shall conclude that a group of low-alpha stars exists at Hercules III and IV. These stars are super metal-rich, super Odd-Z rich, arguably more alpha-poor, and arguably less enhanced in Ba and Li, which indicates older stellar ages.

\section{Discussion} \label{sec:discussion}

The Milky Way galaxy, like many disc galaxies, is found to have a negative radial metallicity gradient in the Galactic plane from the centre to the outer disc \citep[e.g. ][]{Luck2011AJ....142...51L:radialgrad, sanchez2014A&A...570A...6S:radialgrad}. Recent studies of \cite{Wylie2021A&A...653A.143W}, \cite{Lian2023NatAs...7..951L}, and \cite{Guiglion2024A&A...682A...9G} revealed a different, mildly positive [Fe/H] gradient in the inner Galaxy. This results in a $\Lambda$-shaped radial metallicity profile peaked around $R \sim 4-6$\,kpc. In the context of a bar, as stars on bar-like orbits are found more metal-rich \citep{Wegg2019A&A...632A.121W}, this corresponds to the outer region of the Galactic bar, where the super thin bar component is found \citep{Wegg2015MNRAS.450.4050W}. This super metal-rich region, as marked by the yellow circle in Fig.\,\ref{fig:gaiaframe}, is expected to be the result of additional gas inflow and star formation, or capturing of disc stars due to the slowing in bar rotation and increase in bar size. 

The Hercules subgroups, especially the newly included Hercules III and IV, have shown very special chemical trends to field stars and other kinematic groups. Hercules I and II are found to be more Fe enhanced than most field stars (stars not in the kinematic groups) and similar to that of Hyades and Horn in \S\,\ref{sec:X-k}. This feature is consistent with previous analysis by \cite{LiuC2015arXiv151006123L:Herc-Metal-rich,PerezVillegas2017ApJ...840L...2P,Antoja2017A&A...601A..59A,Liang2023ApJ...956..146L}. Moreover, the low alpha thin disc population of Hercules III and IV is found substantially enhanced in iron-peak elements and Odd-Z elements (Na and Al). These structures can be related to HR1614 and Arcturus groups in other studies. Consistent to our analysis, HR1614 is found to be a metal-rich and old kinematic group by \cite{Feltzing2000A&A...357..153F:HR1614,DeSilva2007AJ....133..694D:HR1614,Kushniruk2020A&A...638A.154K}. Moreover, Arcturus or Hercules IV, while believed to be a thick disc or extragalactic stream in early studies \citep[e.g. ][]{Navarro2004ApJ...601L..43N:Arcturus, Bobylev2010AstL...36...27B:Arcturus}, \cite{Kushniruk2019A&A...631A..47K:Arcturus} showed it to be chemically inhomogeneous, but its chemical composition is not shown to be significantly different.

The super-enhancement in iron-peak elements of the low alpha population at Hercules III and IV subgroups found in this paper indicates that these low $L_Z$ Hercules subgroups are not likely to originate locally where the interstellar medium is much less metal-rich. They are more to form closer to the Galactic centre, but due to the positive metallicity gradient in the innermost region of the Galaxy, they are less likely to originate at the Galactic centre. Instead, their place of birth is likely in proximity to the peak of the Galactic metallicity profile, the outer thin bar. Consistent to our analysis, the chemical analysis of many previous studies supported Hercules I and II to have a inner Galactic origin \citep[e.g.]{LiuC2015arXiv151006123L:Herc-Metal-rich, Antoja2017A&A...601A..59A}; the HR1614 group (Hercules III) to have an origin similar to Hercules \citep{Kushniruk2020A&A...638A.154K}; but no agreement has been drawn on the origin of Arcturus except for it not being a dissolved cluster, or a moving group \citep{Monari2013arXiv1306.2632M:arcturus,Kushniruk2019A&A...631A..47K:Arcturus}. In our next paper \citep{LYS2024arXiv241119097L:II}, we propose Trojan orbits, a mechanism that is very likely responsible for Hercules I and II in the slow long bar scenario, to explain the dynamical origin of all four Hercules subgroups.

Moreover, this low $L_Z$ patch of stars is also found to be enhanced in Odd-Z elements Na and Al. Two distinct stellar populations are found in solar-type stars: an old sequence that extends from the high alpha thick disc population to the most Fe-rich low alpha thin disc; and a young sequence with mainly thin disc stars \citep{Ciuca2021MNRAS.503.2814C}. Orbital analysis shows that the old sequence is composed of stars in the inner disc while local and outer disc stars contribute to the young sequence. Moreover, \cite{Nissen2020A&A...640A..81N, Owusu2024arXiv240500315O} showed that enhancement in Na and Al are good indicators of old sequence stars. Therefore, the Na and Al rich stars in Hercules III and IV with deficiency in alpha elements and super enhancement in iron-peak elements are the youngest stars in the old sequence. As the old sequence is originated in the inner disc, this low $L_Z$ patch has an origin in the inner Galaxy.

\begin{figure*}
    \centering
    \includegraphics[width=\linewidth]{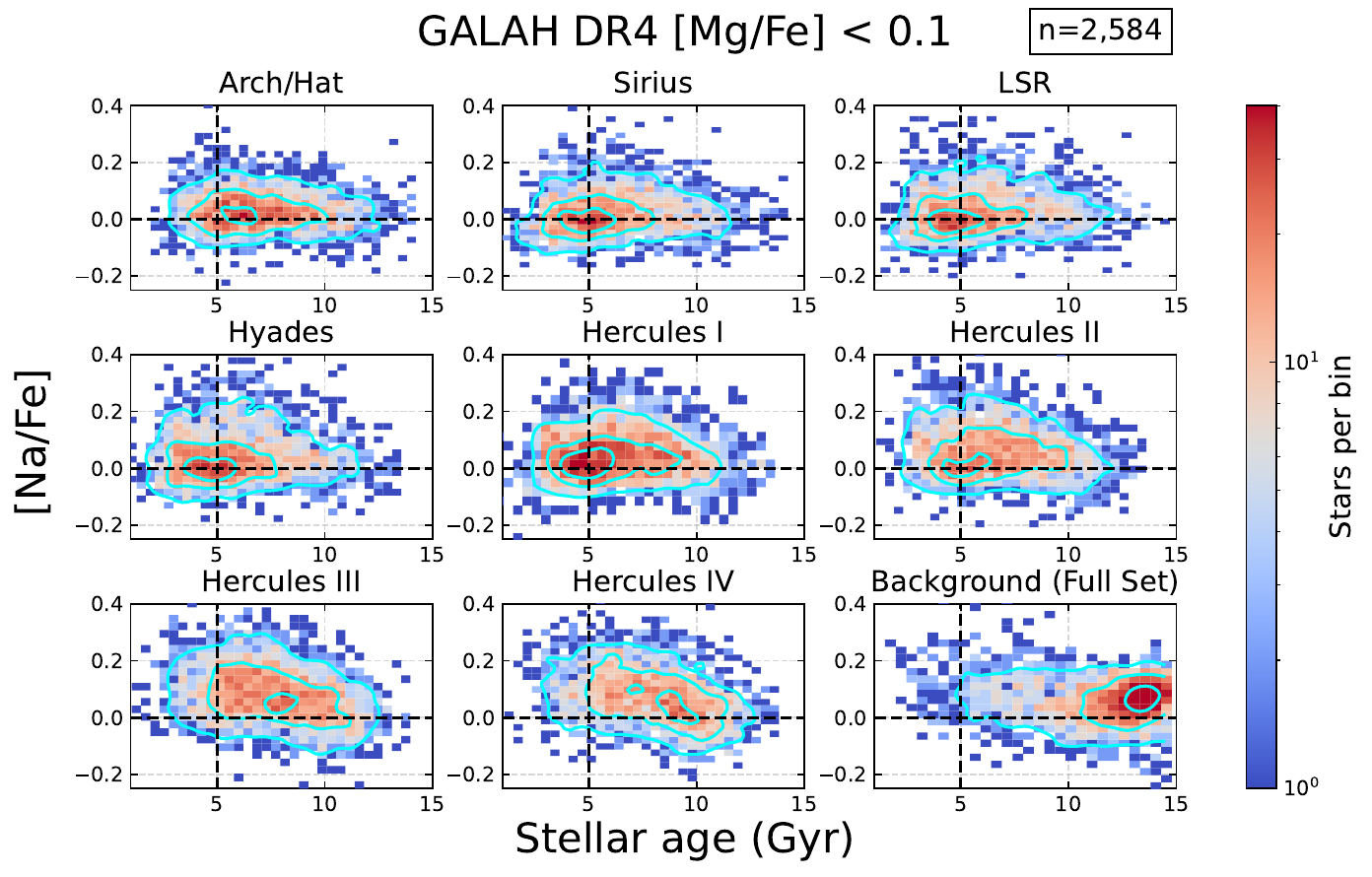}
    \caption{The [Na/Fe]-Age distribution of the SNd GALAH stars in the alpha-deficient kinematic group samples. While the high $L_Z$ kinematic groups occupy similar regions in the diagram, Hercules III and IV include a significant number of intermediate-aged and Na-enhanced stars. The dashed line cross at $(\text{age} = 5\,\text{Gyr},\text{[Na/Fe]}=0.0)$ as the reference. The three cyan iso-density contour lines encircle 90, 60, and 25 per\,cent of stars from the peak of the distribution in each graph. The number of stars in each sample is normalised to $2,584$ by random removal to ensure a constant sample size for better comparison.}
    \label{fig:Na_Age}
\end{figure*}

We extend this discussion with GALAH [Na/Fe] and stellar ages of alpha-deficient MSTO stars. In Fig.\,\ref{fig:Na_Age}, the stellar density distribution in the [Na/Fe]-Age plane is presented in samples of kinematic groups selected from Fig.\,\ref{fig:selection}. The dashed line cross marks $(\text{Age},\text{[Na/Fe]})=(5 \,\text{Gyr},0\,\text{dex})$ as the reference point and the cyan contours encircle 90, 60, and 25 per\,cent of stars from the peak of the distribution just like that in Fig.\,\ref{fig:mgfe_galah}. The kinematic group samples with $L_Z$ higher than Hercules II at the top four panels show a similar distribution that centres around [Na/Fe] $= 0$\,dex and stellar age $\tau \sim 5$\,Gyr. The Arch sample has a peak shifted to a slightly older age and Hercules I and II tend to include slightly more stars with an intermediate age ($\tau \sim 7$\,Gyr) and higher [Na/Fe].

On the other hand, the low $L_Z$ patch at Hercules III and IV shows significantly different distributions. Unlike other kinematic groups that show concentrated distribution centred on the bottom left and nearly flat in the [Na/Fe]-Age plane, Hercules III and IV shows more extended distributions with a higher [Na/Fe] and older ages. The distribution of these stars rises from older stellar ages, $\tau \sim 10$\,Gyr, towards younger ages of $\tau \sim 5$\,Gyr. The peak of this distribution appears around [Na/Fe] $= 0.1$\,dex and stellar age $\tau = 7$\,Gyr. This kinematically selected population is in good agreement with the chemically selected old sequence population in \cite{Owusu2024arXiv240500315O}. By comparing it to the full background sample which is highly concentrated in the oldest Na-enhanced stars peaked at [Na/Fe] $= 0.1$\,dex and stellar age $\tau = 14$\,Gyr, this provides strong support to their identity as the youngest stars in the old sequence.

Furthermore, Li is an interesting fragile element that has an unique evolution track and history. Li is mainly created in the Big Bang nucleosynthesis, often preserved in hotter and faster-rotating stars, and moderated in stellar evolution by depletions in giants and cool stars \citep{Basri1996ApJ...458..600B_Li, EWang2024MNRAS.528.5394W_Li}. Since old stars tend to be less hot, an older population is expected to have less Li because it was already depleted. Therefore, the deficiency in Li in the low $L_Z$ patch suggests a systematically older stellar age. The higher stellar age of this patch is also supported with the deficiency in s-process elements Y and Ba \citep{Nissen2015A&A...579A..52N_sprocess}. The higher age supports the non-local identity as time is required for stars formed inside to travel to the SNd. Furthermore, the Li deficiency in super-solar metallicity in the solar vicinity has been widely studied \citep[e.g.][]{Grisoni2019MNRAS.489.3539G:Li}, and some studies have related their origin to the inner Galaxy \citep{Guiglion2019A&A...623A..99G:Li}, in agreement with our results.

The direct way to verify the origin of the low $L_Z$ Hercules stars is to compare the abundances in the inner disc to that of stars in the SNd. However, as most current optical surveys suffer from dust extinction in the Galactic plane, chemical and kinematical data for stars in the inner disc is scarce. Future infrared surveys into the inner Galaxy may infer more chemical information and thus provide a direct connection between the stellar population in the SNd to that in the inner disc. Based on our analysis, the most useful abundances in addition to Fe and aphas might be Ni, Na, and Al. If the outer bar is found to be Odd-Z rich, the outer bar-origin argument of the super-metal-rich Hercules III and IV will be strongly supported.

With this low $L_Z$ population in Hercules that originates in the inner Galaxy, the next thing to consider is the underlying mechanism that transport these stars out to the SNd. One possible scenario is radial migration. In this scenario, regular circular orbits of stars can be perturbed by the corotation resonance of transient spiral arms, resulting in permanent changes in momenta \citep{Sellwood2002MNRAS.336..785S, Daniel2015MNRAS.447.3576D:RM, Daniel2018MNRAS.476.1561D}. In earlier studies, this has been proposed over the more traditional supercluster scenario \citep{Famaey2005A&A...430..165F:stream}, leading to the concept of dynamical streams (Hercules stream). However, as the radially migrated orbits are still mostly circular, the asymmetry in $V_R$ disfavours the origin of radial migration. Moreover, the significant portion of Hercules stars among the SNd stars ($\sim 23 $\,per\,cent) indicate the possible influence of more stable non-axisymmetric components like the bar. While many previous studies proposed the OLR of the bar as the dynamical origin of Hercules \citep[e.g.][]{Dehnen2000AJ....119..800D, Hunt2018MNRAS.477.3945H_OLR, Fragkoudi2019MNRAS.488.3324F}, the current best bar model rotates slower and is longer \citep{Wegg2015MNRAS.450.4050W, Portail2017MNRAS.465.1621P}. The slower proposed pattern speed pushes the OLR radius beyond the SNd and thus disfavours the OLR scenario. In paper II of this series \citep{LYS2024arXiv241119097L:II}, we will investigate orbits at corotation resonance of the bar, the so called Trojan orbits. We will see the good agreement of this stable mechanism with the chemical signatures shown in this paper, and conclude that the Trojan orbits are an important contributor to all four of the Hercules substructures in the slow-bar scenario \citep[in agreement with][]{PerezVillegas2017ApJ...840L...2P,DOnghia2020ApJ...890..117D,Chiba2021MNRAS.500.4710C_resonace}. Given the current understanding of the inner Galaxy, the Trojan orbits appear as a strong contender for the origin of the Hercules kinematic group.

\section{Conclusions}
\label{sec:summary}

In this paper, we analyse the signatures of Hercules and other kinematic groups in the SNd with chemical data from GALAH DR4 and APOGEE DR17. We arrive on the following conclusions.

\begin{itemize}
    \item We investigated the distribution of chemical abundances in the $L_Z$-$V_R$ kinematic space and found inconsistencies between surveys and inside chemical groups. This anomaly is due to the mixture of Galactic thin and thick disc populations in the sample. We also find that the high [Mg/Fe], low [Fe/H] thick disc stars show limited contribution to the kinematic structures in the SNd and should thus be removed.
    
    \item By removing the high alpha stars, we achieve consistency among elements in the chemical groups. The low $L_Z$ patch of Hercules III and IV stars stands out as an iron-peak enhanced, Odd-Z enhanced, alpha-deficient, and older thin disc population in the SNd. These signatures support their identity as the youngest stars in the old sequence. Moreover, evidence of the existence of such a population is now found consistently in APOGEE and GALAH.
    
    \item The chemical footprints of low alpha Hercules III and IV stars (Fe rich, Na and Al rich, and $\alpha$ poor) suggest their origin in the inner disc. In a barred scenario, they are expected to have an inner Galactic origin around the thin outer bar.
    
    \item Among the low alpha population, stars are chemically similar in Sirius, LSR, Hyades, and Hercules I and II: they are Li rich, Fe rich, alpha poor, and s-process rich young stars. The Arch/hat is relatively alpha rich and Fe-deficient young stars originated from the outer disc. Hercules III and IV are alpha deficient, super Fe-enhanced, Odd-Z enhanced, s-process deficient, and older stars from the inner disc.
\end{itemize}

The Hercules stars originated in the inner Galaxy need to travel into the SNd via some dynamical mechanism. In \cite{LYS2024arXiv241119097L:II}, we will investigate if there is a mechanism that can both form the Hercules $L_Z$-$V_R$ structure in the SNd and is capable of transporting stars from the inner Galaxy into the SNd. We will see that Trojan orbits can explain both traditional high $L_Z$ Hercules I \& II structures and the newly included low $L_Z$ Hecules III \& IV groups. We will also show that these inner disc low $L_Z$ stars are likely to be born on or captured by the fast Trojan quasi-periodic orbits, which can be associated with corotation resonance of the Galactic bar, and pass through the SNd on the orbit.

\section*{Acknowledgements}

We acknowledge the traditional owners of the land on which the AAT and ANU stand, the Gamilaraay, the Ngunnawal and Ngambri people. We pay our respects to elders past, present, and emerging and are proud to continue their tradition of surveying the night sky in the Southern hemisphere.

HJ thanks the Queensland University School of Mathematics and Physics for hospitality and funding support as part of their distinguished visitor program. SB acknowledges support from the Australian Research Council under grant numbers DE240100150. LYS acknowledges friendly discussions with O. Gerhard during his visit to the ANU.

We thank the editors and the anonymous referee for the constructive and straightforward feedback. We acknowledge OUP and MNRAS for kindly granting the APC waiver.

\section*{Software}
The research for this publication was coded in \texttt{Python} v. 3.12 and included packages \texttt{Astropy} v. 6.0.0 \citep{astropy:2013,astropy:2018,astropy:2022}, \texttt{IPython} v. 6.27.1 \citep{ipython:PER-GRA:2007}, \texttt{Matplotlib} v. 3.8.2 \citep{Hunter:2007:matplotlib}, \texttt{NumPy} v. 1.26.4 \citep{harris2020:numpy}, \texttt{SciPy} v. 1.11.4 \citep{2020:SciPy}, and \texttt{tqdm} v. 4.66.1 \citep{tqdm:2022zndo....595120D}.

\section*{Data Availability}
Code used for data analysis and plotting are available on reasonable request. \textit{Gaia} DR3 \citep{Gaia2023A&A...674A...1G}, GALAH DR4 \citep{Buder2024arXiv240919858B}, and APOGEE DR17 \citep{APOGEE2022ApJS..259...35A} can be obtained in their respective websites.

\bibliographystyle{mnras}
\bibliography{bibfile} 

\bsp	
\label{lastpage}
\end{document}